\documentclass[12pt,onecolumn]{article}
\usepackage{amsmath,amssymb,epsfig,psfrag}
\newcommand{\figref}[1]{Fig.~\ref{fig:#1}}
\newcommand{\Real}{{\mathbb{R}}}
\newcommand{\Complex}{{\mathbb{C}}}

\global\def\putFrag#1#2#3#4{
\begin{figure}[tp]
\begin{center}
#4 \epsfxsize=#3in \epsfbox{#1.eps}
\end{center}
\caption{\small{#2}} \label{fig:#1}
\end{figure}
}

\newtheorem{theorem}{Theorem}

\newenvironment{proof}{{\sl Proof\/}:\ \ }%{\qed\vspace{\baselineskip}}

\begin{document}
\title{On the Achievable Diversity-Multiplexing Tradeoffs in Half-Duplex Cooperative Channels}
\author{Kambiz~Azarian, Hesham~El~Gamal, and Philip~Schniter\\
       Dept. of Electrical  and Computer Engineering\\ The Ohio State University\\
       \{azariany,helgamal,schniter\}@ece.osu.edu}
\date{}
\maketitle

\begin{abstract}
\normalsize In this paper, we propose novel cooperative transmission
protocols for delay limited coherent fading channels consisting of
$N$ (half-duplex and single-antenna) partners and one cell site. In
our work, we differentiate between the relay, cooperative broadcast
(down-link), and cooperative multiple-access (up-link) channels. The
proposed protocols are evaluated using Zheng-Tse
diversity-multiplexing tradeoff. For the relay channel, we
investigate two classes of cooperation schemes; namely, Amplify and
Forward (AF) protocols and Decode and Forward (DF) protocols. For
the first class, we establish an upper bound on the achievable
diversity-multiplexing tradeoff with a single relay. We then
construct a new AF protocol that achieves this upper bound. The
proposed algorithm is then extended to the general case with ($N-1$)
relays where it is shown to outperform the space-time coded protocol
of Laneman and Worenell without requiring decoding/encoding at the
relays. For the class of DF protocols, we develop a dynamic decode
and forward (DDF) protocol that achieves the optimal tradeoff for
multiplexing gains $0\leq r\leq 1/N$. Furthermore, with a single
relay, the DDF protocol is shown to dominate the class of AF
protocols for all multiplexing gains. The superiority of the DDF
protocol is shown to be more significant in the cooperative
broadcast channel. The situation is reversed in the cooperative
multiple-access channel where we propose a new AF protocol that
achieves the optimal tradeoff for all multiplexing gains. A
distinguishing feature of the proposed protocols in the three
scenarios is that they do not rely on orthogonal subspaces, allowing
for a more efficient use of resources.  In fact, using our results
one can argue that the sub-optimality of previously proposed
protocols stems from their use of orthogonal subspaces rather than
the half-duplex constraint.
\end{abstract}
%\spacing{1.5}
\section{Introduction}

%\subsection{Motivation}
Recently, there has been a growing interest in the design and
analysis of wireless cooperative transmission protocols (e.g.,
\cite{SEA:031}-\cite{ZME:04}). These works consider several
interesting scenarios (e.g., fading-vs-AWGN channels,
ergodic-vs-quasistatic channels, and full-duplex-vs-half-duplex
transmission) and devise appropriate transmission techniques and
analysis tools, based on the settings. Here, we focus on the
delay-limited coherent channel and adopt the same setup as
considered by Laneman, Tse, and Wornell in \cite{LTW:02}. There, the
authors imposed the half-duplex constraint (either transmit or
receive, but not both) on the cooperating nodes and proposed several
cooperative transmission protocols. In this setup, the basic idea is
to leverage the antennas available at the other nodes in the network
as a source of \emph{virtual} spatial diversity. The proposed
protocols in \cite{LTW:02} were classified as either Amplify and
Forward (AF), where the helping node retransmits a scaled version of
its soft observation, or Decode and Forward (DF), where the helping
node attempts first to decode the information stream and then
re-encodes it using (a possibly different) code-book. All the
proposed schemes in \cite{LTW:02} used a Time Division Multiple
Access (TDMA) strategy, where the two partners relied on the use of
orthogonal subspaces to repeat each other's signals. Later, Laneman
and Wornell extended their DF strategy to the $N$ partners scenario
\cite{LW:03}. Other follow-up works have focused on developing
practical coding schemes that attempt to exploit the promised
information theoretic gains (e.g., \cite{SE:03,JHHN:04}).

As observed in \cite{LTW:02, LW:03}, previously proposed
cooperation protocols suffer from a significant loss of
performance in high spectral efficiency scenarios. In fact, the
authors of \cite{LTW:02} posed the following open problem: {\em
``a key area of further research is exploring cooperative
diversity protocols in the high spectral efficiency regime.''}
This remark motivates our work here, where we present more
efficient (and in some cases optimal) AF and DF protocols for the
relay, cooperative broadcast (CB), and cooperative multiple-access
(CMA) channels. To establish the gain offered by the proposed
protocols, we adopt the diversity-multiplexing tradeoff as our
measure of performance. This powerful tool was introduced by Zheng
and Tse for point-to-point multi-input-multi-output (MIMO)
channels in \cite{ZT02} and later used by Tse, Viswanath, and
Zheng to study the (non-cooperative) multiple-access channel in
\cite{TVZ:03}.

In the following, we summarize the main results of this paper, some
of which were initially reported in
\cite{AES:03,AES:04,E:04,AES:042,AES:043}.

\begin{enumerate}
\item For the single relay channel, we establish an upper bound on
the achievable diversity-multiplexing tradeoff by the class of AF
protocols. We then identify a variant within this class, referred to
as the Nonorthogonal Amplify and Forward (NAF) protocol, that
achieves this upper bound.  We then propose a dynamic decode and
forward (DDF) protocol and show that it achieves the \emph{optimal}
tradeoff for multiplexing gains $0\leq r\leq 0.5$\footnote{The
multiplexing gain ``$r$'' will be defined rigorously in the sequel}.
Furthermore, the DDF protocol is shown to outperform all AF
protocols for arbitrary multiplexing gains. Finally, the two
protocols (i.e., NAF and DDF) are extended to the scenario with
$N-1$ relays where we characterize their tradeoff curves. Notably,
the NAF protocol is shown to outperform the space-time coded
protocol of Laneman and Wornell (LW-STC) \cite{LW:03} without
requiring decoding/encoding at the relays.

\item For the cooperative broadcast channel, we present a modified
version of the DDF protocol to allow for reliable transmission of
the common information. We then characterize the tradeoff curve of
this protocol and use this characterization to establish its
superiority compared to AF protocols. In fact, we argue that the
gain offered by the DDF is more significant in this scenario (as
compared to the relay channel).

\item For the symmetric multiple-access scenario, we propose a
novel AF cooperative protocol where an {\em artificial}
inter-symbol-interference (ISI) channel is created. We prove the
optimality (in the sense of diversity-multiplexing tradeoff) of this
protocol by showing that, for all multiplexing gains (i.e., $0 \leq
r \leq 1$), it achieves the diversity-multiplexing tradeoff of the
corresponding $N\times 1$ point-to-point channel. One can then use
this result to argue that the sub-optimality of the schemes proposed
in \cite{LTW:02} was dictated by the use of orthogonal subspaces
rather than the half-duplex constraint. We also utilize this result
to shed more light on the fundamental difference between half-duplex
relay and cooperative multiple-access channels.
\end{enumerate}

Before proceeding further, a brief remark regarding two independent
parallel works \cite{NBK:04,PV:04} is in order. In \cite{NBK:04},
Nabar, Bolcskei and Kneubuhler considered the half-duplex
single-relay channel, under \emph{almost} the same assumptions as in
\cite{LTW:02} (i.e., the only difference is that, for diversity
analysis, the relay-destination channel was assumed to be
non-fading) and proposed a set of AF and DF protocols. In one of
their AF protocols (NBK-AF), Nabar \emph{et. al.} allowed the source
to continue transmission over the whole duration of the codeword,
while the relay listened to the source for the first half of the
codeword and relayed the received signal over the second half. This
makes the NBK-AF protocol identical to the NAF protocol proposed in
this paper. Here, we characterize the diversity-multiplexing
tradeoff achieved by this protocol while relaxing the assumption of
non-fading relay-destination channel. Using this analysis, we
establish the optimality of this scheme within the class of linear
AF protocols. Furthermore, we generalize the NAF protocol to the
case of arbitrary number of relays and characterize its achieved
tradeoff curve. In \cite{PV:04}, Prasad and Varanasi derived upper
bounds on the diversity-multiplexing tradeoffs achieved by the DF
protocols proposed in \cite{NBK:04}. In the sequel, we establish the
gain offered by the proposed DDF protocol by comparing its
diversity-multiplexing tradeoff with the upper bounds in
\cite{PV:04}. Finally, we emphasize that, except for the
single-relay NAF protocol, all the other protocols proposed in this
paper are novel.

In this paper, we use $(x)^+$ to mean $\max\{x,0\}$, $(x)^-$ to mean
$\min\{x,0\}$ and $\lceil x \rceil$ to mean nearest integer to $x$
towards plus infinity. $\Real^{N}$ and $\Complex^{N}$ denote the set
of real and complex $N$-tuples, respectively, while $\Real^{N+}$
denotes the set of non-negative $N$-tuples. We denote the complement
of set $O \subseteq \Real^{N}$, in $\Real^{N}$, by $O^c$, while
$O^+$ means $O \cap \Real^{N}$. $I_{N}$ denotes the $N\times N$
identity matrix, $\Sigma_{\mathbf{x}}$ denotes the autocovariance
matrix of vector $\mathbf{x}$, and $\log(.)$ denotes the base-$2$
logarithm.

The rest of the paper is organized as follows. In
Section~\ref{back}, we detail our modeling assumptions and review,
briefly, some results that will be extensively used in the sequel.
The half-duplex relay channel is investigated in
Section~\ref{relay} where we describe the NAF and DDF protocols
and derive their tradeoff curves. In Section~\ref{broad}, we
extend the DDF protocol to the cooperative broadcast channel.
Section~\ref{mac} is devoted to the cooperative multiple-access
channel where we propose a new AF protocol and establish its
optimality, in the symmetric scenario, with respect to the
diversity-multiplexing tradeoff. In Section~\ref{num}, we present
numerical results that show the SNR gains offered by the proposed
schemes in certain representative scenarios. Finally, we offer
some concluding remarks in Section~\ref{conc}. To enhance the flow
of the paper, we collect all the proofs in the Appendix.

%\subsection{Outline}

%\subsection{Notation}

\section{Background}\label{back}
First we state the general assumptions that apply to the three
scenarios considered in this paper (i.e., relay, broadcast, and
multiple-access). Assumptions pertaining to a specific scenario
will be given in the related section.

\begin{enumerate}
\item All channels are assumed to be flat Rayleigh-fading and
quasi-static, i.e., the channel gains remain constant during a
coherence-interval and change independently from one
coherence-interval to another. Furthermore, the channel gains are
mutually independent with unit variance. The additive noises at
different nodes are zero-mean, mutually-independent,
circularly-symmetric and white complex-Gaussian. Furthermore, the
variances of these noises are proportional to one another such that
there will always be \emph{fixed} offsets between the different
channels' signal to noise ratios (SNRs).

\item All nodes have the same power constraint, have a single
antenna, and operate synchronously. Only the receiving node of any
link knows the channel gain; no feedback to the transmitting node
is permitted (the incremental relaying protocol proposed in
\cite{LTW:02} can not, therefore, be considered in our framework).
Following in the footsteps of \cite{LTW:02}, all cooperating
partners operate in the half-duplex mode, i.e., at any point in
time, a node can either transmit or receive, but not both. This
constraint is motivated by, e.g., the typically large difference
between the incoming and outgoing signal power levels. Though this
half-duplex constraint is quite restrictive to protocol
development, it is nevertheless assumed throughout the paper.

\item Throughout the paper, we assume the use of random Gaussian
code-books where a codeword spans the entire coherence-interval of
the channel. Furthermore, we assume asymptotically large
code-lengthes. This implies that the diversity-multiplexing
tradeoffs derived in this paper, serve as upper-bounds for the
performance of the proposed protocols with finite code-lengths.
Results related to the design of practical coding/decoding schemes
that approach the fundamental limits established here will be
reported elsewhere.
\end{enumerate}

Next we summarize several important definitions and results that
will be used throughout the paper.
\begin{enumerate}
\item The SNR of a link, $\rho$, is defined as
\begin{align}
\rho &\triangleq \frac{E}{\sigma_v^2}, \label{eq:44}
\end{align}
where $E$ denotes the average energy available for transmission of a
symbol across the link and $\sigma_v^2$ denotes the variance of the
noise observed at the receiving end of the link. We say that
$f(\rho)$ is \emph{exponentially equal to} $\rho^b$, denoted by
$f(\rho) \dot{=} \rho^b$, when
\begin{align}
\lim_{\rho \rightarrow \infty} \frac{\log (f(\rho))}{\log (\rho)} &=
b. \label{eq:45}
\end{align}
In \eqref{eq:45}, $b$ is called the \emph{exponential order} of $f(\rho)$.
$\dot{\leq}$ and
$\dot{\geq}$ are defined similarly.

\item Consider a family of codes $\{C_{\rho}\}$ indexed by
operating SNR $\rho$, such that the code $C_{\rho}$ has a rate of
$R(\rho)$ bits per channel use (BPCU) and a maximum likelihood (ML)
error probability $P_E(\rho)$. For this family, the
\emph{multiplexing gain} ``$r$'' and the \emph{diversity gain}
``$d$'' are defined as
\begin{align}
&r \triangleq \lim_{\rho\rightarrow\infty}\frac{R(\rho)}{\log\rho} ,
&d \triangleq
-\lim_{\rho\rightarrow\infty}\frac{\log(P_E(\rho))}{\log\rho} .
\label{eq:1}
\end{align}

\item The problem of characterizing the optimal tradeoff between the
reliability and throughput of a point-to-point communication system
over a coherent quasi-static flat Rayleigh-fading channel was posed
and solved by Zheng and Tse in \cite{ZT02}. For a MIMO communication
system with $M$ transmit and $N$ receive antennas, they showed that,
for any $r \leq \min\{M,N\}$, the optimal diversity gain $d^*(r)$ is
given by the piecewise linear function joining the $(r,d)$ pairs
$(k,(M-k)(N-k))$ for $k=0,...,\min\{M,N\}$, provided that the
code-length $l$ satisfies $l\ge M+N-1$.

\item We say that protocol $A$ \emph{uniformly dominates} protocol $B$ if,
for any multiplexing gain $r$, $d_A(r) \geq d_B(r)$.

\item Assume that $g$ is a Gaussian random variable with zero mean and
unit variance. If $v$ denotes the exponential order of $1/|g|^2$, i.e.,
\begin{align}
v &= -\lim_{\rho \rightarrow \infty}
\frac{\log(|g|^2)}{\log(\rho)}, \label{eq:46}
\end{align}
then the probability density function (PDF) of $v$ can be shown to
be:
\begin{align}
p_v &= \lim_{\rho \rightarrow \infty} \ln(\rho) \rho^{-v}
\exp(-\rho^{-v}) . \nonumber
\end{align}
Careful examination of the previous expression reveals that
\begin{align}
p_{v} &\dot{=}
\begin{cases}
\rho^{-\infty}= 0, &\text{~for~} v<0, \\
\rho^{-v}, &\text{~for~} v\geq 0
\end{cases}. \label{eq:47}
\end{align}
Thus, for independent random variables $\{v_j\}_{j=1}^{N}$
distributed identically to $v$, the probability $P_O$ that $(v_1,
\dots, v_N)$ belongs to set $O$ can be characterized by
\begin{align}
P_O &\dot{=} \rho^{-d_o} \text{~~for~~} d_o =\inf_{(v_1, \dots, v_N)
\in O^+} \sum_{j=1}^N v_j, \label{eq:48}
\end{align}
provided that $O^+$ is not empty. In other words, the exponential
order of $P_O$ only depends on $O^+$. This is due to the fact that
the probability of any set, consisting of $N$-tuples
$(v_1,\dots,v_N)$ with at least one negative element, decreases
exponentially with SNR and therefore can be neglected compared to
$P_{O^+}$ which decreases polynomially with SNR.

\item Consider a coherent linear Gaussian channel where a random
Gaussian code-book is used. The pairwise error probability (PEP) of
the ML decoder, denoted as $P_{PE}$, averaged over the ensemble of
random Gaussian codes, is upper bounded by
\begin{align}
P_{PE} &\leq \det(I_N + \frac{1}{2} \Sigma_{\mathbf{s}}
\Sigma_{\mathbf{n}}^{-1})^{-1}, \label{eq:49}
\end{align}
where $\mathbf{s}\in\Complex^N$ and $\mathbf{n}\in\Complex^N$
denote the signal and noise components of the observed vector,
respectively (i.e., $\mathbf{y} = \mathbf{s} + \mathbf{n}$).
\end{enumerate}

\section{The Half-Duplex Relay Channel}\label{relay}
In this section, we consider the relay scenario in which $N-1$
relays help a single source to better transmit its message to the
destination. As the vague descriptions ``help'' and ``better
transmit'' suggest, the general relay problem is rather broad and
only certain sub-problems have been studied (for example see
\cite{CE:79}). In this work, we focus on two important classes of
relay protocols. The first is the class of Amplify and Forward (AF)
protocols, where a relaying node can only process the observed
signal linearly before re-transmitting it. The second is the class
of Decode and Forward (DF) protocols, where the relays are allowed
to decode and re-encode the message using (a possibly different)
code-book. Here we emphasize that, a priori, it is not clear which
class (i.e., AF or DF) offers a better performance (e.g.,
\cite{LTW:02}).

\subsection{Amplify and Forward Protocols}
We first consider the single relay scenario (i.e., $N=2$). For
this scenario, we derive the optimal diversity-multiplexing
tradeoff and identify a specific protocol within this class, i.e.,
the NAF protocol, that achieves this optimal tradeoff. We then
extend the NAF protocol to the general case with an arbitrary
number of relays.

Under the half-duplex constraint, it is easy to see that any
single-relay AF protocol can be mathematically described by some
choice of the matrices $A_1$, $A_2$, and $B$ in the following
model
\begin{align}
\mathbf{y} &= \begin{bmatrix} g_1A_{1} & 0 \\ g_2hBA_1 & g_1A_2
\end{bmatrix} \mathbf{x} + \begin{bmatrix} 0 \\ g_2B \end{bmatrix}
\mathbf{w} + \mathbf{v}. \label{eq:51}
\end{align}
In \eqref{eq:51}, $\mathbf{y}\in\Complex^{l}$ represents the vector
of observations at the destination, $\mathbf{x}\in\Complex^{l}$ the
vector of source symbols, $\mathbf{w}\in\Complex^{l'}$ the vector of
noise samples (of variance $\sigma_w^2$) observed by the relay, and
$\mathbf{v}\in\Complex^{l}$ the vector of noise samples (of variance
$\sigma_v^2$) observed by the destination. The variables $h$, $g_1$
and $g_2$ denote the source-relay channel gain, source-destination
channel gain, and relay-destination channel gain, respectively.
$A_{1}\in\Complex^{l'\times l'}$ and
$A_{2}\in\Complex^{(l-l')\times(l-l')}$ are diagonal matrices. In
this protocol, the source can potentially transmit a new symbol in
every symbol-interval of the codeword, while the relay listens
during the first $l'$ symbols and then, for the remaining $l-l'$
symbols, transmits linear combinations of the $l'$ noisy
observations using the coefficients in $B\in\Complex^{(l-l')\times
l'}$. In fact, by letting $l'=l/2$, $A_1=I_{l'}$, $A_2=0$ and
$B=bI_{l'}$ (with $b \leq \sqrt{E/(|h|^2E+\sigma^2_w)}$ denoting the
relay repetition gain), we obtain Laneman-Tse-Wornell Amplify and
Forward (LTW-AF) protocol \cite{LTW:02}. Finally, we note that when
the source symbols are independent, the average energy constraint
translates to
\begin{align}
|h|^2E\sum_{i=1}^{l'}|b_{ji}|^2|a_i|^2+\sigma_w^2\sum_{i=1}^{l'}|b_{ji}|^2
&\leq E, \quad j=1, \dots, l-l' , \label{eq:28}
\end{align}
where $B=[b_{ji}]$ and $A_1=\text{diag}(a_1,\cdots,a_{l'})$.

\begin{theorem} \label{thrm:1}
The optimal diversity gain for the cooperative relay scenario with
a single AF relay is upper-bounded by
\begin{align}
d^*(r) &\leq (1-r)+(1-2r)^+. \label{eq:36}
\end{align}
\end{theorem}
\begin{proof}
Please refer to the Appendix.
\end{proof}

%\subsubsection{Single Relay NAF Protocol}

The upper-bound on $d^*(r)$, as given by \eqref{eq:36}, is shown
in \figref{d_r_naf}. Having Theorem~\ref{thrm:1} at hand, it now
suffices to identify an AF protocol that achieves this upper-bound
in order to establish its optimality. Towards this end, we observe
that, in the proof of Theorem~\ref{thrm:1}, the only requirements
on $B$ such that the protocol described by \eqref{eq:51} could
\emph{potentially} achieve the optimal diversity-multiplexing
tradeoff are for $B$ to be square (of dimension $l/2 \times l/2$)
and full-rank. Furthermore, $B$ should not violate the relay
average energy constraint as given by \eqref{eq:28}. Thus, the
simple choices
\begin{align}
A_1 &= I_{l/2} &A_2 &= I_{l/2} &B &= bI_{l/2} \text{~~for~~} b \leq
\sqrt{ \frac{E}{|h|^2E+\sigma_w^2}} \label{eq:54}
\end{align}
inspire our NAF protocol. In particular, the source transmits on
every symbol-interval in a cooperation frame, where a cooperation
frame is defined as two consecutive symbol-intervals. The relay,
on the other hand, transmits only once per cooperation frame; it
simply repeats the (noisy) signal it observed during the previous
symbol-interval. It is important to realize that this design is
dictated by the half-duplex constraint, which implies that the
relay can repeat at most once per cooperation frame. We denote the
repetition gain by $b$ and, for frame $k$, we denote the
information symbols by $\{x_{j,k}\}_{j=1}^{2}$. The signals
received by the destination during frame $k$ are thus:
\begin{align}
y_{1,k}&=g_1x_{1,k}+v_{1,k} \nonumber \\
y_{2,k}&=g_1x_{2,k}+g_2b(hx_{1,k}+w_{1,k})+v_{2,k} \nonumber
\end{align}
where the repetition gain $b$ must satisfy \eqref{eq:54}. Note that,
in order to decode the message, the destination needs to know the
relay repetition gain $b$, the source-relay channel gain $h$, the
source-destination channel gain $g_1$, and the relay-destination
channel gain $g_2$. Now, we are ready to establish the optimality of
the NAF protocol with respect to the diversity-multiplexing
tradeoff.

%\subsubsection{Single Relay NAF Diversity-Multiplexing Tradeoff}
\begin{theorem} \label{thrm:2}
The NAF protocol achieves the optimal diversity-multiplexing
tradeoff for the AF single-relay scenario, which is:
\begin{align}
d^*(r) &= (1-r)+(1-2r)^+. \label{eq:55}
\end{align}
\end{theorem}
\begin{proof}
Please refer to the Appendix.
\end{proof}

Three remarks are now in order:
\begin{enumerate}
\item As shown in \figref{d_r_naf}, the NAF protocol enjoys
uniform dominance over the direct transmission scheme (i.e., no
cooperation) and LTW-AF protocol. This dominance can be attributed
to relaxing the orthogonality constraint whereby one can reap two
distinct benefits: rate enhancement via continuous transmission and
diversity enhancement via cooperation. It is interesting to note
that this dominance is achieved while only half of the symbols are
repeated by the relay.

\item From \figref{d_r_naf}, one can see that for multiplexing
gains greater that $0.5$, the diversity gain achieved by the
proposed NAF relay protocol is identical to that of the
non-cooperative protocol. This is due to the fact that the
\emph{AF cooperative} link provided by the relay can not support
multiplexing gains greater than $0.5$---a consequence of the
half-duplex constraint. Hence, for multiplexing gains larger than
$0.5$, there is only one link from the source to the destination,
and, thus, the tradeoff curve is identical to that of a
point-to-point system with one transmit and one receive antenna.
Later, we will show that the proposed DDF strategy avoids this
drawback.

\item As shown in the proof of Theorem~\ref{thrm:2}, the
achievability of the optimal tradeoff is not very sensitive to the
choice of the repetition gain ``$b$'' (i.e., for a wide range of
choices, the NAF protocol achieves the optimal tradeoff). In
practice, one should optimize the repetition gain, experimentally
if needed, to minimize the outage probability at the target rate
and signal-to-noise ratio.
\end{enumerate}
%\subsubsection{NAF For $N-1$ Relay Nodes}
The NAF protocol can be extended to the case of arbitrary number of
relays (i.e., $N\geq 2$) as follows. First, we define a super-frame
as a concatenation of $N-1$ consecutive cooperation frames. Within
each super-frame, the relays take turns repeating the signals they
previously observed as they did in the case of a single relay (refer
to \figref{multi-relay_naf}). Thus, the destination's received
signals during a super-frame will be
\begin{align}
y_{1,1}&=g_1x_{1,1}+v_{1,1} \nonumber \\
y_{2,1}&=g_1x_{2,1}+g_2b_2(h_2x_{1,1}+w_{1,1})+v_{2,1} \nonumber \\
y_{1,2}&=g_1x_{1,2}+v_{1,2} \nonumber \\
y_{2,2}&=g_1x_{2,2}+g_3b_3(h_3x_{1,2}+w_{1,2})+v_{2,2} \nonumber \\
\vdots & \nonumber \\
y_{1,N-1}&=g_1x_{1,N-1}+v_{1,N-1} \nonumber \\
y_{2,N-1}&=g_1x_{2,N-1}+g_Nb_N(h_Nx_{1,N-1}+w_{1,N-1})+v_{2,N-1} ,
\nonumber
\end{align}
where the source-relay channel gain, relay-destination channel gain,
relay repetition gain, and relay-observed noise for relay
$i\in\{1,...,N-1\}$ are denoted by $h_{i+1}$, $g_{i+1}$, $b_{i+1}$
and $w_{1,i}$, respectively. As before, $g_1$ represents the
source-destination channel gain. The quantities $y_{j,k}$,
$v_{j,k}$, and $x_{j,k}$ represent the received signal, noise
sample, and source symbol, respectively, during the $j^{\text{th}}$
symbol-interval of the $k^{\text{th}}$ cooperation frame. Note that
there is nothing to be gained by having more than one relay
transmitting the same symbol simultaneously. Also, similar to the
single-relay NAF scenario, the destination needs to know all relay
repetition gains $\{b_i\}_{i=2}^N$ as well as all channel gains
$\{g_i\}_{i=1}^N$ and $\{h_i\}_{i=2}^N$. The following Theorem
characterizes the diversity-multiplexing tradeoff achieved by this
protocol.
\begin{theorem} \label{thrm:3}
The diversity-multiplexing tradeoff achieved by the NAF protocol
with $N-1$ relays is characterized by
\begin{align}
d(r) &= (1-r)+(N-1)(1-2r)^+ . \nonumber
\end{align}
\end{theorem}
\begin{proof}The proof is virtually identical to that of Theorem~\ref{thrm:2}, and hence, is omitted for brevity.
\end{proof}

It is interesting to note that the generalized NAF protocol
uniformly dominates the LW-STC. This can be attributed to the fact
that in the generalized NAF protocol, in contrast to the LW-STC
protocol, the source transmits over the whole duration of the
codeword. The generalized NAF protocol offers the additional
advantage of low complexity since it does not require
decoding/encoding at the relays.

\subsection{Decode and Forward Protocols}
In this class of protocols, we allow for the possibility of
decoding/encoding at the different relays. In \cite{LTW:02},
Laneman-Tse-Wornell presented a particular variant of DF protocols
(LTW-DF) where the source transmits in the first half of the
codeword. Based on its received signal in this interval, the relay
attempts to decode the message. It then re-encodes and transmits the
encoded stream in the second half of the codeword. In \cite{LW:03},
Laneman and Wornell derived the diversity-multiplexing tradeoff
achieved by this scheme (i.e., $d(r)= 2(1-2r)$), which is depicted
in \figref{d_r_ddf_2}. Here, we propose a Dynamic Decode and Forward
(DDF) protocol and characterize its tradeoff curve. This
characterization reveals the uniform dominance of this protocol over
all known \emph{full-diversity} (i.e., $d(0)=2$) protocols proposed
for the half-duplex single-relay channel and furthermore establishes
its optimality, over a certain range of multiplexing gains (i.e., $0
\leq r \leq 1/2$). We first describe and analyze the protocol for
the case of a single relay. Generalization to $N-1$ relays will then
follow.

Similar to the previous section, we assume that a codeword consists
of $l$ consecutive symbol-intervals, during which all the channel
gains remain unchanged. In the DDF protocol, the source transmits
data at a rate of $R$ BPCU during every symbol-interval in the
codeword. The relay, on the other hand, listens to the source until
the mutual information between its received signal and source signal
exceeds $lR$. It then decodes and re-encodes the message using an
\emph{independent} Gaussian code-book and transmits it during the
rest of the codeword. The dynamic nature of the protocol is
manifested in the fact that we allow the relay to listen for a time
duration that depends on the instantaneous channel realization to
maximize the probability of successful decoding. We denote the
signals transmitted by the source and relay as $\{x_k\}_{k=1}^l$ and
$\{\tilde{x}_k\}_{k=l'+1}^{l}$, respectively, where $l'$ is the
number of symbol-intervals the relay waits before starting
transmission. Using this notation, the received signals (at the
destination) can be written as:
\begin{align}
\begin{array}{llc}
y_k=g_1x_k+v_k                 &\text{for} &l' \geq k \geq 1, \\
y_k=g_1x_k+g_2\tilde{x}_k+v_k  &\text{for} &l \geq k > l'.
\end{array} \nonumber
\end{align}
From the protocol description, it is clear that the number of
symbols where the relay listens should be chosen as:
\begin{align}
l' &=
\min\left\{l,\left\lceil\frac{lR}{\log_2{(1+|h|^2c\rho)}}\right\rceil\right\},
\label{eq:61}
\end{align}
where $h$ is the source-relay channel gain, and
$c=\sigma_v^2/\sigma_w^2$. One can now see the dependence of this
choice of $l'$ on the instantaneous channel realization and that
this choice, together with the asymptotically large $l$, guarantees
that when $l'<l$, the relay average probability of error with a
Gaussian code ensemble is arbitrarily small\footnote{This point will
be established rigorously in the proof of Theorem \ref{thrm:4}}.
Clearly, when $l'=l$ the relay does not contribute to the
transmission of the message, and hence, incorrect decoding at the
relay in this case does not affect performance. Here, we observe
that, in contrast to the NAF protocol, the destination does not need
to know the source-relay channel gain. It does, however, need to
know the relay waiting time $l'$, along, with the source-destination
and relay-destination channel gains. The following Theorem describes
the diversity-multiplexing tradeoff achievable with this cooperation
protocol.

%\subsubsection{Single Relay DDF Diversity-Multiplexing Tradeoff}
\begin{theorem} \label{thrm:4}
The diversity-multiplexing tradeoff achieved by the single relay DDF
protocol is given by
\begin{align}
d(r) &= \left\{
\begin{array}{lll}
2(1-r) & \text{if} & \frac{1}{2} \geq r \geq 0 \\
(1-r)/r & \text{if} & 1 \geq r \geq \frac{1}{2}
\end{array} \label{eq:10}
\right. .
\end{align}
\end{theorem}
\begin{proof}
Please refer to the Appendix.
\end{proof}

The diversity-multiplexing tradeoff of \eqref{eq:10} is shown in
\figref{d_r_ddf_2}. It is now clear that the DDF protocol is optimal
for $0\leq r\leq 0.5$ since it achieves the genie aided diversity
(where the relay is assumed to know the information message {\em
a-priori}). For $r> 0.5$, the DDF protocol suffers from a loss,
compared to the genie aided strategy, since, on the average, the
relay will only be able to help during a small fraction of the
codeword. It is easy to see that, the performance for this range of
multiplexing gains can not be improved through employing a mixed AF
and DF strategy. In fact, the DDF strategy dominates all such
strategies\footnote{The proof for this is rather straightforward,
and hence, is omitted here for brevity.}. It remains to be seen
whether there exists a strategy that closes the gap to the genie
aided strategy when $r>0.5$ or not. Note also that the gain offered
by the DDF protocol, compared to AF protocols, can be attributed to
the ability of this strategy to transmit independent Gaussian
symbols after successful decoding. In AF strategies, on the other
hand, the relay is limited to repeating the noisy Gaussian symbols
it receives from the source. \figref{d_r_ddf_2} also compares the
DDF protocol with the DF protocol proposed in \cite{NBK:04}, which
we refer to as NBK-DF. In this comparison, we utilize the
upper-bound derived by Prasad and Varanasi on the
diversity-multiplexing tradeoff of the NBK-DF, which was reported in
\cite{PV:04}. One can see from \figref{d_r_ddf_2} that the NBK-DF
protocol does not achieve any diversity gain greater than one. This
can be attributed to the fact that in this protocol, the message is
split up into two parts, out of which, only one is retransmitted by
the relay. \figref{d_r_ddf_2} also shows that for multiplexing gains
close to one, the NBK-DF upper-bound outperforms the DDF protocol.
Therefore, in this range, the comparison between the two protocols
depends on the tightness of the NBK-DF upper-bound which was not
discussed in \cite{PV:04}.

%\subsubsection{DDF For $N-1$ Relay Nodes}
Next, we describe the generalization of the DDF protocol to the case
of multiple relays. In this case, the source and relays cooperate in
nearly the same manner as in the single relay case. Specifically,
the source transmits during the whole codeword while each relay
listens until the mutual information between its received signal and
the signals transmitted by the source and other relays exceeds $lR$.
It is assumed that every relay knows the code-books used by the
source and other relays. Once a relay decodes the message, it uses
an independent code-book to re-encode the message, which it then
transmits for the rest of the codeword. Note that, since the
source-relay channel gains may differ, the relays may require
different wait times for decoding. This complicates the protocol,
since a given relay's ability to decode the message requires precise
knowledge of the times at which every other relay begins its
transmission. To address this problem, the codeword is divided into
a number of segments, and relays are allowed to start transmission
only at the beginning of a segment. In between the segments, every
relay is allowed to broadcast a (well protected) beacon, informing
all other relays whether or not it will start transmission.
Judicious choice of the segment length, relative to the codeword
length, results in only a small loss compared to the genie-aided
case, whereby all relays know all decoding times {\em a-priori}.
Here, we assume that the number of segments is sufficiently large
and the length of the beacon signals is much smaller than the
segment length. Therefore, in characterizing the
diversity-multiplexing tradeoff achieved by this protocol, we ignore
the losses associated with the beacons and the quantization of the
starting times for the different relays.
\begin{theorem} \label{thrm:5}
The diversity-multiplexing tradeoff achieved by the DDF protocol
with $N-1$ relays is characterized by:
\begin{align}
d(r) &= \begin{cases}
N(1-r), & \frac{1}{N} \geq r \geq 0, \\
1+\frac{(N-1)(1-2r)}{1-r}, & \frac{1}{2} \geq r \geq \frac{1}{N}, \\
\frac{1-r}{r}, & 1 \geq r \geq \frac{1}{2}.
\end{cases} \label{eq:37}
\end{align}
\end{theorem}
\begin{proof}
Please refer to the Appendix.
\end{proof}

The diversity-multiplexing tradeoff \eqref{eq:37} is shown in
\figref{d_r_ddf_5} and \figref{d_r_ddf_diff} for different values
of $N$. While the loss of the DDF protocol compared to the
genie-aided protocol increases with $N$, it is not clear at the
moment if this loss is due to the half-duplex constraint or due to
the sub-optimality of the DDF strategy.

\section{The Half-Duplex Cooperative Broadcast Channel}\label{broad}

We now consider the cooperative broadcast (CB) scenario, where a
single source broadcasts to $N$ destinations. The destinations
cooperate through helping one another in receiving their messages.
We assume that the source message for destination $j \in
\{1,\cdots,N\}$ consists of two parts. A common part of rate $R_c =
r_c\log(\rho)$ BPCU, which is intended for all of the destinations
and an individual part of rate $R_j=r_j\log(\rho)$ BPCU, which is
specific to the $j^{th}$ destination. The total rate is then
$R=R_c+\sum_{j=1}^N R_j$ and the multiplexing gain tuple is given by
$\mathbf{r}=\left(r_c,r_1,...,r_N\right)$. We define the overall
diversity gain $d$ based on the performance of the worst receiver as
\begin{align}
d &= \min_{1\leq j\leq N}\{d_j\}, \nonumber
\end{align}
where we require all the receivers to decode the common
information\footnote{Clearly this definition does not allow for
different Quality of Service (QoS) constraints.}. Now, as a first
step, one can see that if $r_c=0$, i.e. if there is no common
message, then the techniques developed for the relay channel can be
{\em exported} to this setting through a proportional time sharing
strategy. With this assumption, all of the properties of the NAF and
DDF protocols, established for the relay channel, carry over to this
scenario. The problem becomes slightly more challenging when
$r_c>0$. In fact, it is easy to see that, for a fixed total rate
$R$, the highest probability of error corresponds to the case where
all destinations are required to decode all the messages. This
translates to the following condition (that applies to any
cooperation scheme)

\begin{equation} d(r_c,r_1,r_2,...,r_N)\geq
d(r_c+r_1+...+r_N,0,0,..,0).
\end{equation}
So, we will focus the following discussion on this worst case
scenario, i.e.,

\begin{equation}
\mathbf{r}=\left(r_c,0,0,...,0\right), 0\leq r_c\leq
1.
\end{equation}

The first observation is that, in this scenario, the only AF
strategy that achieves the full rate extreme point $( r=1, d=0)$ is
the non-cooperative protocol. Any other AF strategy will require
some of the nodes to re-transmit, and therefore not to {\em listen}
during parts of the codeword\footnote{This follows from the
half-duplex constraint.}, which prevents it from achieving full
rate. Fortunately, this drawback can be avoided in the DDF protocol.
The reason is that, in this protocol, any node will start helping
only after it has successfully decoded the message.
%\subsection{CB-DDF Protocol}
We now propose a protocol for the CB scenario that is a direct
extension of the DDF relay protocol. This will be referred to as the
CB-DDF protocol in the sequel. The only modification needed,
compared to the relay channel case, is that now every destination
can act as a relay for the other destinations, based on its
instantaneous channel gain. Specifically, the source transmits
during the whole codeword while each destination listens until the
mutual information between its received signal and the signals
transmitted by the source and other destinations exceeds $lR$. Once
a destination decodes the message, it uses an independent code-book
to re-encode the message, which it then transmits for the rest of
the codeword. Similar to the relay channel, it is assumed that every
destination knows the code-books used by the source and other
destinations. Also, the protocol must include a mechanism that keeps
every destination informed of the re-transmission starting times of
all the other destinations. Again, in deriving the following result,
we ignore the associated cost of this mechanism, relying on the
asymptotic assumptions.
\begin{theorem} \label{thrm:6}
The diversity-multiplexing tradeoff achieved by the CB-DDF
protocol with $N$ destinations is given by:
\begin{align}
d(r_c) &= \begin{cases}
N(1-r_c), & \frac{1}{N} \geq r_c \geq 0, \\
1+\frac{(N-1)(1-2r_c)}{1-r_c}, & \frac{1}{2} \geq r_c \geq \frac{1}{N}, \\
\frac{1-r_c}{r_c}, & 1 \geq r_c \geq \frac{1}{2}.
\end{cases} \label{eq:41}
\end{align}
\end{theorem}
\begin{proof}
Please refer to the Appendix.
\end{proof}

It is interesting to note that this is exactly the same tradeoff
obtained in the relay channel. This implies that requiring all nodes
to decode the message does not entail a price in terms of the
achievable tradeoff.
\section{The Half-Duplex Cooperative Multiple-Access Channel}\label{mac}

In this section, we consider the cooperative multiple-access (CMA)
scenario, where $N$ sources transmit their independent messages to a
common destination. We assume symmetry so that all sources transmit
information at the same rate and are limited by the same power
constraint. The basic idea of the proposed protocol, which we refer
to as the CMA-NAF protocol, is to create an artificial ISI channel.
Towards this end, each of the $N$ sources transmits once per
cooperation frame, where a cooperation frame is defined as $N$
consecutive symbol-intervals (refer to part a of \figref{cma_naf}).
Each source is assigned unique {\em transmission and reception}
symbol-intervals within the cooperation frame. During its
transmission symbol-interval, a source transmits a linear
combination of its own symbol and the signal it observed during its
most recent reception symbol-interval. In other words, every source,
in addition to sending its own symbol, \emph{helps} another source
by repeating the (noisy) signal it last received from it. Without
loss of generality, we set the $j^{th}$ source transmission
symbol-interval equal to $j$.

We now provide an illustrative example for the $N=3$ case. Here we
assume that sources $1$, $2$, and $3$ help sources $3$, $1$, and
$2$, respectively. For the $j^{th}$ source and the $k^{th}$
cooperation frame, $t_{j,k}$ denotes the transmission, $r_{j,k}$
the (assigned) reception, and $x_{j,k}$ the originating symbol.
Using $a_j$ and $b_j$ to denote the broadcast and repetition gains
of the $j^{th}$ source, respectively, the signals transmitted
during the first two cooperation frames would be (in chronological
order)
\begin{align}
t_{1,1}&=a_1x_{1,1} \nonumber \\
t_{2,1}&=a_2x_{2,1}+b_2r_{2,1} \nonumber \\
t_{3,1}&=a_3x_{3,1}+b_3r_{3,1} \nonumber \\
t_{1,2}&=a_1x_{1,2}+b_1r_{1,1} \nonumber \\
t_{2,2}&=a_2x_{2,2}+b_2r_{2,2} \nonumber \\
t_{3,2}&=a_3x_{3,2}+b_3r_{3,2} . \nonumber
\end{align}
Using $h_{ji}$ to denote the $i^{th}$-source-to-$j^{th}$-source
channel gain, and $w_{j,k}$ to denote the noise observed by the
$j^{th}$ source during its $k^{th}$-frame reception
symbol-interval, the assigned receptions become
\begin{align}
r_{2,1}&=h_{21}t_{1,1}+w_{2,1} \nonumber \\
r_{3,1}&=h_{32}t_{2,1}+w_{3,1} \nonumber \\
r_{1,1}&=h_{13}t_{3,1}+w_{1,1} \nonumber \\
r_{2,2}&=h_{21}t_{1,2}+w_{2,2} \nonumber \\
r_{3,2}&=h_{32}t_{2,2}+w_{3,2}. \nonumber
\end{align}
Using $g_j$ to denote the $j^{th}$-source-to-destination channel
gain, and $v_{j,k}$ to denote the noise observed by the destination
during the $j^{th}$ symbol-interval of the $k^{th}$ frame, the
signals observed at the destination would be
\begin{align}
y_{j,k}&=g_jt_{j,k}+v_{j,k} \nonumber.
\end{align}
The source-observed noises $\{w_{j,k}\}$ have variance
$\sigma_w^2$ for all $j,k$, and the destination-observed noises
$\{v_{j,k}\}$ have variance $\sigma_v^2$ for all $j,k$. Note that,
as mandated by our half-duplex constraint, no source transmits and
receives simultaneously. The broadcast and repetition gains
$\{a_j,b_j\}$ should be chosen to satisfy the average power
constraint
\begin{align}
E\{|t_{j,k}|^2\} &\le E . \label{eq:19}
\end{align}

Let us now define $L$ consecutive cooperation frames as a
super-frame (refer to part b of \figref{cma_naf}). We will assume
that helper assignments are fixed within a super-frame but are
scheduled to change across super-frames. We impose the following
requirements on helper scheduling.
\begin{enumerate}
\item In each super-frame, every source is helped by a different source.
\item Across super-frames, every source is helped equally by every other source.
\end{enumerate}
Among the many scheduling rules that satisfy these requirements, we
choose the following circular rule. In super-frame $i$, sources with
indices $(1,\dots,N)$ are assigned helpers with indices given by the
$j^{th}$ right circular shift of $(1,\dots,N)$, where $j=\langle i-1
\rangle_{N-1} + 1$. For example, when $N=4$, the helper
configurations are given by the following table.
\begin{center}
\begin{tabular}{|c|@{~~~~}c@{~~~~}c@{~~~~}c@{~}c|}
\hline
Super-frame & \multicolumn{4}{l|}{\hspace{-1.5em}Helper assigned to}\\
index & 1 & 2 & 3 & 4 \\
\hline
1 & 4 & 1 & 2 & 3 \\
2 & 3 & 4 & 1 & 2 \\
3 & 2 & 3 & 4 & 1 \\
4 & 4 & 1 & 2 & 3 \\
\hline
\end{tabular}
\end{center}
Since this scheduling algorithm generates $N-1$ distinct helper
configurations, the length of the super-frames, $L$, is chosen such
that a coherence-interval consists of $N-1$ consecutive super-frames
(refer to part c of \figref{cma_naf}). To achieve maximal diversity
for a given multiplexing gain, it is required that all codewords
span the entire coherence-interval. For this reason, we choose codes
of length $l$ given by
\begin{align}
l &= (N-1)L . \label{eq:15}
\end{align}

%\subsection{CMA-NAF Diversity-Multiplexing Tradeoff}
Similar to the broadcast channel, defining the multiplexing gain
$r$ and diversity gain $d$ for the cooperative multiple-access
channel requires some care. Note that, using \eqref{eq:1}, the
pair $(r_j,d_j)$ can be defined for communication between the
$j^{th}$ source and the destination. However, since we assumed a
symmetric CMA setup, all multiplexing gains are equal, i.e.,
$r=r_j$ for all $j$. Furthermore, since CMA-NAF mandates that only
one source transmits in any symbol-interval, the destination's
multiplexing gain is also equal to $r$. That is, the destination
receives information at rate $R$ given by
\begin{align}
R &= r\log(\rho). \label{eq:58}
\end{align}
We define the overall diversity gain $d$ based on the worst case
probability of error for the $N$ information streams, i.e.,
\begin{align}
d &= \min_{1 \leq j \leq N}\{d_j\} . \nonumber
\end{align}
With these definitions, Theorem \ref{thrm:7} establishes the
optimality of the CMA-NAF in the symmetric scenario with $N$
sources.
\begin{theorem} \label{thrm:7}
The CMA-NAF protocol achieves the optimal (genie-aided)
diversity-multiplexing tradeoff for the symmetric scenario with $N$
sources, given by
\begin{align}
d^*(r) &= N(1-r) . \label{eq:14}
\end{align}
\end{theorem}
\begin{proof}
Please refer to the Appendix.
\end{proof}

Theorem~\ref{thrm:7} not only establishes the optimality of the
CMA-NAF protocol, but also it shows that the half-duplex constraint
does not entail any cost, in terms of diversity-multiplexing
tradeoff, in the symmetric CMA channel. One can now attribute the
sub-optimality of the CMA schemes reported in \cite{LTW:02,LW:03} to
the use of orthogonal subspaces. It is interesting to observe that
one can achieve the optimal tradeoff in the symmetric CMA channel
with a simple AF strategy. In fact, by comparing
Theorems~\ref{thrm:1}~and~\ref{thrm:7}, one can see the fundamental
difference between the half duplex CMA and relay channels.

\section{Numerical Results}\label{num}
In this section, we report numerical results that quantify the
performance gains offered by the proposed protocols. These numerical
results correspond to outage probabilities and are meant to show
that the superiority of the proposed protocols in terms of
diversity-multiplexing tradeoff translates into significant SNR
gains. In \figref{out_naf}, \figref{out_ddf}, and
\figref{out_cma_naf}, we compare the proposed protocols with the
non-cooperative (direct transmission) and the LTW-AF protocols. To
ensure fairness, we have imposed more strict power constraints on
the NAF and the DDF relay protocols; specifically, we lowered the
average transmission energy of the source and the relay from $E$ to
$E/2$ during the interval when both are transmitting. This way, the
total average energy per symbol-interval, spent by any of the
protocols considered here\footnote{In the CMA-NAF protocol, the
constant average energy per symbol interval is automatically
implied.} is $E$. While one may find other energy allocation
strategies that offer performance improvement (in terms of the
outage probability), any such optimization will not affect the
achievable diversity-multiplexing tradeoff, and hence, will not be
pursued here. To obtain lower bounds on the gains offered by the DDF
and CMA-NAF protocols, we assume a noiseless source-relay channel
for the LTW-AF and NAF relay protocols. For the DDF relay and the
CMA-NAF protocols, the SNR of the link between the two cooperating
partners was assumed to be only 3 dB better than that of the
relay-destination or source-destination channels. We optimized the
broadcast and repetition gains for the CMA-NAF protocol
experimentally. In all the considered cases, the outage
probabilities are computed through Monte-Carlo simulations.

\figref{out_naf} shows the performance gain offered by the NAF relay
protocol over both the non-cooperative protocol and the LTW-AF
protocol at high SNRs and two different data rates. The same
comparison is repeated in \figref{out_ddf} with the DDF protocol
where, as expected, the gains are shown to be larger. The CMA
channel is considered in \figref{out_cma_naf} where the optimality
of the CMA-NAF protocol is shown to translate into significant SNR
gains. It is also interesting to note that the gap between CMA-NAF
performance and genie-aided strategy is less than $3$ dB when the
date rate is equal to 2 BPCU. We can also observe that the gains
offered by the DDF and CMA-NAF protocols compared with the LTW-AF
protocol increase with the data rate. This is a direct consequence
of the higher multiplexing gains achievable with our newly proposed
protocols. Overall, these results re-emphasize the fact that the
full diversity criterion alone\footnote{Full diversity corresponds
to the point $(d=2, r=0)$ on the tradeoff curve.} is a rather weak
design tool.

We conclude this section with a brief comment on our choice for the
diversity-multiplexing tradeoff as our design tool. This choice is
inspired by the convenient tradeoff, between analytical tractability
and accuracy, that this tool offers. Ideally, one should seek
cooperation schemes that minimize the outage probability at the
target rate and SNR. Unfortunately, it is easy to see that such an
approach would lead to an intractable problem even in very
simplified scenarios. Our results, on the other hand, demonstrate
that one can use the diversity-multiplexing tradeoff to analytically
guide the design in many relevant scenarios. From the accuracy point
of view, our simulation results validate that schemes with better
tradeoff characteristics {\em always} offer significant SNR gains at
sufficiently high SNRs. In this context, the main drawback of the
diversity-multiplexing tradeoff is that it fails to predict at which
SNR the promised gains will start to appear. For example, from the
figures, one can see that the DDF and CMA-NAF schemes yield
performance gains at relatively moderate SNRs whereas the NAF
protocol only offers gain at larger SNRs.

\section{Conclusions}\label{conc}
In this paper, we considered the design of cooperative protocols for
a system consisting of half-duplex nodes. In particular, we
differentiated between three scenarios. For the relay channel, we
investigated  the AF and DF protocols. We established the uniform
dominance of the proposed DDF protocol compared to all known full
diversity cooperation strategies and its optimality in a certain
range of multiplexing gains. We then proceeded to the cooperative
broadcast channel where the gain offered by the DDF strategy was
argued to be more significant, as compared to the relay channel. For
the multiple-access scenario, we proposed a novel AF cooperative
protocol where an {\em artificial} ISI channel was created. We
proved the optimality (in the sense of diversity-multiplexing
tradeoff) of this protocol by showing that it achieves the same
tradeoff curve as the genie-aided $N\times 1$ point-to-point system.

Our results reveal interesting insights on the structure of optimal
cooperation strategies with half-duplex partners. First, we observe
that, without the half-duplex constraint, achieving the optimal
tradeoff in the three channels considered here is rather
straightforward (i.e., one can easily construct a simple AF strategy
that results in an $N$-tap ISI channel, and hence, the optimal
tradeoff). With the half-duplex constraint, more care is necessary
in constructing the cooperation strategies, but, as shown, one can
still achieve the optimal tradeoff in many relevant scenarios. One
of the important insights is that one should strive to transmit {\em
independent} symbols as frequently as possible. Indeed, the
optimality of the proposed CMA-NAF protocol stems from exploiting
the distributed nature of the information to enable transmission of
an independent symbol in every symbol interval. It is now easy to
see that the use of orthogonal subspaces to enable cooperation, as
in \cite{LTW:02} for example, entails a significant loss in the
achievable tradeoff.

This work poses many interesting questions. For example, proving (or
disproving) the optimality of the DDF protocol for the single relay
channel and $r>0.5$ is an open problem. Generalizations of the
proposed schemes to multi-antenna nodes, cooperative Automatic
Retransmission reQuest (ARQ) channels \cite{AES:042}, scenarios with
different QoS constraints, and asymmetric CMA channels are of
definite interest. Finally, the design of practical coding/decoding
strategies that approach the fundamental limits achievable with
Gaussian codes and maximum likelihood decoding is an important venue
to pursue.

\section{Appendix}
In this section, we collect all the proofs.

\subsection{Proof of Theorem~\ref{thrm:1}}
Due to the source average energy constraint, setting $A_1$ and
$A_2$ to anything other than the identity matrix will reduce the
mutual information between $\mathbf{x}$ and $\mathbf{y}$. Since we
are interested in obtaining an upper bound, we will choose
$A_1=I_{l'}$ and $A_2=I_{l-l'}$, in which case \eqref{eq:51}
reduces to
\begin{align}
\mathbf{y} &= \begin{bmatrix} g_1I_{l'} & 0 \\ g_2hB & g_1I_{l-l'}
\end{bmatrix} \mathbf{x} + \begin{bmatrix} 0 \\ g_2B \end{bmatrix}
\mathbf{w} + \mathbf{v} . \label{eq:27}
\end{align}
Using singular value decomposition (SVD), the matrix $B$ can be
factored as
\begin{align}
B &= UDV^H , \nonumber
\end{align}
where $U\in\Complex^{(l-l') \times (l-l')}$ and $V\in\Complex^{l'
\times l'}$ are unitary and where $D\in\Complex^{(l-l') \times l'}$
is non-negative diagonal with the diagonal elements in decreasing
order. Using these matrices, we define $\tilde{\mathbf{y}}
\triangleq T\mathbf{y}$, $\tilde{\mathbf{x}} \triangleq
T\mathbf{x}$, $\tilde{\mathbf{v}} \triangleq T\mathbf{v}$, and
$\tilde{\mathbf{w}} \triangleq V^H\mathbf{w}$, for unitary
transformation
\begin{align}
T &\triangleq \begin{bmatrix} V^H & 0 \\
0 & U^H
\end{bmatrix} . \nonumber
\end{align}
The unitary property of $V$ and $T$ implies that
$\Sigma_{\tilde{\mathbf{w}}}=\sigma_w^2I_{l}$ and
$\Sigma_{\tilde{\mathbf{v}}}=\sigma_v^2I_{l}$, as well as
\begin{align}
I(\mathbf{x};\mathbf{y}) &=
I(\tilde{\mathbf{x}};\tilde{\mathbf{y}}). \label{eq:33}
\end{align}
In terms of the new variables, \eqref{eq:27} becomes
\begin{align}
\tilde{\mathbf{y}} &= \begin{bmatrix} g_1I_{l'} & 0 \\
g_2hD & g_1I_{l-l'}
\end{bmatrix} \tilde{\mathbf{x}} + \begin{bmatrix} 0 \\ g_2D
\end{bmatrix} \tilde{\mathbf{w}} + \tilde{\mathbf{v}} \nonumber \\
&= \begin{bmatrix} g_1I_{l'} & 0 \\
g_2hD & g_1I_{l-l'} \end{bmatrix} \tilde{\mathbf{x}} +
\tilde{\mathbf{n}} \label{eq:29}
\end{align}
with
\begin{align}
\Sigma_{\tilde{\mathbf{n}}} &=
\begin{bmatrix} \sigma_v^2I_{l'} & 0 \\
0 & \sigma_v^2I_{l-l'}+|g_2|^2 \sigma_w^2DD^H \end{bmatrix} .
\nonumber
\end{align}
If we denote the non-zero diagonal elements of $D$ as
$\{d_i\}_{i=1}^m$, then \eqref{eq:29} can be written as
\begin{align}
\tilde{\mathbf{y}}_i &= G_i \tilde{\mathbf{x}}_i +
\tilde{\mathbf{n}}_i,\quad i = 1, \dots, m \nonumber \\
\tilde{y}_i &= g_1 \tilde{x}_i + \tilde{n}_i, \quad i = m+1, \dots,
l' \text{~and~} i=l'+m+1, \dots, l, \nonumber
\end{align}
where $\tilde{y}_i$, $\tilde{x}_i$ and $\tilde{n}_i$ represent the
$i^{\text{th}}$ element of $\tilde{\mathbf{y}}$,
$\tilde{\mathbf{x}}$ and $\tilde{\mathbf{n}}$, respectively, and
where $\tilde{\mathbf{y}}_i \triangleq [\tilde{y}_i,
\tilde{y}_{l'+i}]^t$, $\tilde{\mathbf{x}}_i \triangleq
[\tilde{x}_i, \tilde{x}_{l'+i}]^t$, $\tilde{\mathbf{n}}_i
\triangleq [\tilde{n}_i, \tilde{n}_{l'+i}]^t$, and
\begin{align}
G_i &\triangleq \begin{bmatrix} g_1 & 0 \\ g_2hd_i & g_1
\end{bmatrix},
    \label{eq:107} \\
\Sigma_{\tilde{\mathbf{n}}_i} &= \begin{bmatrix} \sigma_v^2 & 0 \\
0 & \sigma_v^2+|g_2|^2d_i^2\sigma_w^2 \end{bmatrix} . \label{eq:108}
\end{align}
Note that, according to the SVD theorem,
\begin{align}
m &\leq \min\{l',l-l'\} . \label{eq:30}
\end{align}
Because $\Sigma_{\tilde{\mathbf{n}}}$ is diagonal,
$I(\tilde{\mathbf{x}};\tilde{\mathbf{y}})$ (and therefore
$I(\mathbf{x};\mathbf{y})$) is maximized when $\{
\tilde{\mathbf{x}}_i\}_{i = 1}^{m} \cup
\{\tilde{x}_i\}_{i=m+1}^{l'} \cup \{\tilde{x}_i\}_{i=l'+m+1}^{l}$
are mutually independent, in which case we would have
\begin{align}
\max_{\Sigma_{\tilde{\mathbf{x}}}}I(\tilde{\mathbf{x}};\tilde{\mathbf{y}})
&= \sum_{i=1}^m
\max_{\Sigma_{\tilde{\mathbf{x}}_i}}I(\tilde{\mathbf{x}}_i;\tilde{\mathbf{y}}_i)
+ \sum_{i=m+1}^{l'} \max I(\tilde{x}_i;\tilde{y}_i) +
\sum_{i=l'+m+1}^{l} \max I(\tilde{x}_i;\tilde{y}_i) . \label{eq:31}
\end{align}
The mutual information between $\tilde{\mathbf{x}}_i$ and
$\tilde{\mathbf{y}}_i$ is given by
\begin{align}
I(\tilde{\mathbf{x}}_i;\tilde{\mathbf{y}}_i) &= \log(\det{(I_2 +
\Sigma_{\tilde{\mathbf{n}}_i}^{-\frac{1}{2}} G_i
\Sigma_{\tilde{\mathbf{x}}_i} G_i^H
\Sigma_{\tilde{\mathbf{n}}_i}^{-\frac{1}{2}})}) . \label{eq:32}
\end{align}
A lower-bound on $\max_{\Sigma_{\tilde{\mathbf{x}}_i}}
I(\tilde{\mathbf{x}}_i;\tilde{\mathbf{y}}_i)$ is easily obtained by
replacing $\Sigma_{\tilde{\mathbf{x}}_i}$ by $EI_2$:
\begin{align}
\log(\det{(I_2 + EG_iG_i^H \Sigma_{\tilde{\mathbf{n}}_i}^{-1})})
&\leq \max_{\Sigma_{\tilde{\mathbf{x}}_i}}
I(\tilde{\mathbf{x}}_i;\tilde{\mathbf{y}}_i) . \label{eq:105}
\end{align}
Since $\log(\det{(.)})$ is an increasing function on the cone of
positive-definite Hermitian matrices and since $\lambda_{\text{max}}
I_2-\Sigma_{\tilde{\mathbf{x}}_i} \ge 0$ (where
$\lambda_{\text{max}}$ represents the largest eigenvalue of
$\Sigma_{\tilde{\mathbf{x}}}$), we get the following upper-bound on
$\max_{\Sigma_{\tilde{\mathbf{x}}_i}}
I(\tilde{\mathbf{x}}_i;\tilde{\mathbf{y}}_i)$:
\begin{align}
\max_{\Sigma_{\tilde{\mathbf{x}}_i}} I(\tilde{\mathbf{x}}_i ;
\tilde{\mathbf{y}}_i) &\le \log(\det{(I_2 + \lambda_{\text{max}}
G_iG_i^H \Sigma_{\tilde{\mathbf{n}}_i}^{-1})}) . \label{eq:106}
\end{align}
From \eqref{eq:105} and \eqref{eq:106}, we conclude that
\begin{align}
\frac{\log(\det{(I_2 + EG_iG_i^H
\Sigma_{\tilde{\mathbf{n}}_i}^{-1})})}{\log(\rho)} &\leq
\frac{\max_{\Sigma_{\tilde{\mathbf{x}}_i}} I(\tilde{\mathbf{x}}_i ;
\tilde{\mathbf{y}}_i)}{\log(\rho)} \leq \frac{\log(\det{(I_2 +
\lambda_{\text{max}} G_iG_i^H
\Sigma_{\tilde{\mathbf{n}}_i}^{-1})})}{\log(\rho)}.\nonumber
\end{align}
Now, since $\lambda_{\text{max}}$ is of the same exponential order
as $E$, the bounds converge as $\rho$ grows to infinity. That is
\begin{align}
\lim_{\rho \rightarrow \infty}
\frac{\max_{\Sigma_{\tilde{\mathbf{x}}_i}, d_i}
I(\tilde{\mathbf{x}}_i ; \tilde{\mathbf{y}}_i)}{\log(\rho)} &=
\lim_{\rho \rightarrow \infty} \frac{\log(\det{(I_2 + EG_iG_i^H
\Sigma_{\tilde{\mathbf{n}}_i}^{-1})})}{\log(\rho)}.\nonumber
\end{align}
Plugging in for $G_i$ and $\Sigma_{\tilde{\mathbf{n}}_i}$ from
\eqref{eq:107} and \eqref{eq:108}, respectively, we get
\begin{align}
\lim_{\rho \rightarrow \infty}
\frac{\max_{\Sigma_{\tilde{\mathbf{x}}_i}, d_i}
I(\tilde{\mathbf{x}}_i ; \tilde{\mathbf{y}}_i)}{\log(\rho)} =&
\lim_{\rho \rightarrow \infty} \frac{1}{\log(\rho)}
\log(1+\frac{|g_1|^2E}{\sigma_v^2} + \cdots \nonumber \\
&\frac{(|g_1|^2+|g_2|^2|h|^2|d_i|^2)E}{\sigma_v^2+|g_2|^2d_i^2\sigma_w^2}+
\frac{|g_1|^4E^2}{\sigma_v^2(\sigma_v^2+|g_2|^2d_i^2\sigma_w^2)})
.\nonumber
\end{align}
It is then straightforward to see that
\begin{align}
\lim_{\rho \rightarrow \infty}
\frac{\max_{\Sigma_{\tilde{\mathbf{x}}_i}, d_i}
I(\tilde{\mathbf{x}}_i ; \tilde{\mathbf{y}}_i)}{\log(\rho)} &=
(\max\{2(1-v_1), 1-(v_2+u)\})^+, \label{eq:109}
\end{align}
where $v_1, v_2$ and $u$ are the exponential orders of $1/|g_1|^2$,
$1/|g_2|^2$ and $1/|h|^2$, respectively. In deriving this
expression, we have assumed that $(v_1,v_2,u) \in \Real^{3+}$; as
explained earlier, we do not need to consider realizations in which
$v_1$, $v_2$ or $u$ are negative. Similarly,
\begin{align}
\lim_{\rho \rightarrow \infty} \frac{\max I(\tilde{x}_i ;
\tilde{y}_i)}{\log(\rho)} &= (1-v_1)^+, \nonumber
\end{align}
which, together with \eqref{eq:109}, \eqref{eq:33} and
\eqref{eq:31}, results in:
\begin{align}
\lim_{\rho \rightarrow \infty} \frac{\max_{\Sigma_{\mathbf{x}}}
I(\mathbf{x} ; \mathbf{y})}{\log(\rho)} &=
(l-2m)(1-v_1)^++m(\max\{2(1-v_1), 1-(v_2+u)\})^+. \label{eq:34}
\end{align}
For the quasi-static fading setup, the outage event is defined as
the set of channel realizations for which the instantaneous
capacity falls below the target data rate. Thus, our outage event
$O$ becomes
\begin{align}
O &= \{(v_1,v_2,u)| \max_{\Sigma_{\mathbf{x}}}
I(\mathbf{x},\mathbf{y}) < lR \} . \nonumber
\end{align}
Letting $R$ grow with $\rho$ according to
\begin{align}
R &= r\log(\rho) , \nonumber
\end{align}
and using \eqref{eq:34}, we conclude that, for large $\rho$,
\begin{align}
O^+ = \{(v_1,v_2,u) \in \Real^{3+} | &(l-2m)(1-v_1)^+ + \cdots
\nonumber \\ &m(\max\{2(1-v_1), 1-(v_2+u)\})^+ < rl \} ,
\label{eq:35}
\end{align}
and thus
\begin{align}
P_O(R) &\dot{=} \rho^{-d_o(r)}\text{~~for~~} d_o(r) =
\inf_{(v_1,v_2,u) \in O^+} (v_1+v_2+u) . \label{eq:52}
\end{align}
As Zheng and Tse have shown in Lemma 5 of \cite{ZT02}, $d_o(r)$
provides an upper-bound on $d^*(r)$ (i.e., the optimal diversity
gain at multiplexing gain $r$):
\begin{align}
d^*(r) &\leq d_o(r) . \label{eq:53}
\end{align}
From \eqref{eq:35} and \eqref{eq:52}, it is easy to see that the
right hand side of \eqref{eq:53} is maximized when $m$ is set to its
maximum, which, according to \eqref{eq:30}, is $\min\{l', l-l'\}$.
This is the case when $B$ is full-rank. On the other hand, $\min
\{l', l-l'\}$ itself is maximized when $l'=l/2$ (assuming an even
codeword length $l$), which corresponds to $B$ being a square
matrix. For this $B$, $d_o(r)$ can be shown to take the value of the
right hand side of \eqref{eq:36}.  This completes the proof.

\subsection{Proof of Theorem~\ref{thrm:2}}
The proof closely follows that for the MIMO point-to point
communication system in \cite{ZT02}. In particular, we assume that
the source uses a Gaussian random code-book of codeword length $l$,
where $l$ is taken to be even, and data rate $R$, where $R$
increases with $\rho$ according to
\begin{align}
R &= r\log(\rho) . \nonumber
\end{align}
The error probability of the ML decoder, $P_E(\rho)$, can be upper
bounded using Bayes' rule:
\begin{align}
P_E(\rho) &= P_O(R)P_{E|O}+P_{E,O^c} \nonumber \\
P_E(\rho) &\leq P_O(R)+P_{E,O^c} , \nonumber
\end{align}
where $O$ denotes the outage event. The outage event $O$ is chosen
such that $P_O(R)$ dominates $P_{E,O^c}$, i.e.,
\begin{align}
P_{E,O^c} &\dot{\leq} P_O(R) , \label{eq:4}
\end{align}
in which case
\begin{align}
P_E(\rho) &\dot{\leq} P_O(R) . \label{eq:9}
\end{align}
In order to characterize $O$, we note that, since the destination
observations during different frames are independent, the
upper-bound on the ML conditional PEP [recalling \eqref{eq:49}],
assuming $l$ to be even, changes to
\begin{align}
P_{PE|g_1,g_2,h} &\leq
\det\left(I_{2}+\frac{1}{2}\Sigma_s\Sigma_n^{-1}\right)^{-l/2} ,
\label{eq:57}
\end{align}
where $\Sigma_{\mathbf{s}}$ and $\Sigma_{\mathbf{n}}$ denote the
covariance matrices of destination observation's signal and noise
components during a single frame:
\begin{align}
\Sigma_{\mathbf{s}} &= \begin{bmatrix} |g_1|^2 & g_1g_2^*b^*h^* \\
g_1^*g_2bh & |g_1|^2+|g_2|^2|bh|^2 \end{bmatrix}E  \label{eq:56b} \\
\Sigma_{\mathbf{n}} &=
\begin{bmatrix} \sigma_v^2 & 0 \\ 0 & \sigma_v^2+
|g_2|^2|b|^2\sigma_w^2 \end{bmatrix} . \label{eq:56}
\end{align}
Let us define $v_1$, $v_2$, $u$, and $w$ as the exponential orders
of $1/|g_1|^2$, $1/|g_2|^2$, $1/|h|^2$, and $|b|^2$, respectively.
Then the constraint on $b$ given in \eqref{eq:54} implies the
following constraint on $w$:
\begin{align}
w &\leq \min\{u,1\} \label{eq:5}
\end{align}
We assume $b$ is chosen such that the exponential order $w$
becomes
\begin{align}
w\triangleq(u)^- . \nonumber
\end{align}
which satisfies the constraint given by \eqref{eq:5}.
Interestingly, if we consider $(v_1,v_2,u) \in \Real^{3+}$, then
$w$ becomes zero and vanishes in the expressions. Plugging
\eqref{eq:56b}-\eqref{eq:56} into \eqref{eq:57}, we obtain
\begin{align}
P_{PE|v_1,v_2,u} &\dot{\leq} \rho^{-\frac{l}{2}(\max\{2(1-v_1),
1-(v_2+u)\})^+} \text{~~for~~} (v_1,v_2,u) \in \Real^{3+} .
\nonumber
\end{align}
With rate $R=r\log{\rho}$ BPCU and codeword length $l$, we have a
total of $\rho^{rl}$ codewords. Thus,
\begin{align}
P_{E|v_1,v_2,u} &\dot{\leq} \rho^{-\frac{l}{2}[(\max\{2(1-v_1),
1-(v_2+u)\})^+-2r]} \text{~~for~~} (v_1,v_2,u) \in \Real^{3+}  .
\nonumber
\end{align}
$P_{E,O^c}$ is the average of $P_{E|v_1,v_2,u}$ over the set of
channel realizations that do not cause an outage (i.e., $O^c$).
Using \eqref{eq:47}, one can see that
\begin{align}
P_{E,O^c} &\dot{\leq}
\int_{O^{c+}}\rho^{-d_e(r,v_1,v_2,u)}dv_1dv_2du . \nonumber
\end{align}
for
\begin{align}
d_e(r,v_1,v_2,u) &= \frac{l}{2}[(\max\{2(1-v_1), 1-(v_2+u)
\})^+-2r]+(v_1+v_2+u) . \nonumber
\end{align}
Now, $P_{E,O^c}$ is dominated by the term corresponding to the
minimum value of $d_e(r,v_1,v_2,u)$ over $O^{c+}$:
\begin{align}
P_{E,O^c} &\dot{\leq} \rho^{-d_e(r)} \text{~~for~~} d_e(r) =
\inf_{v_1,v_2,u\in O^{c+}}d_e(r,v_1,v_2,u) . \label{eq:7}
\end{align}
Using \eqref{eq:48}, $P_O(R)$ can be expressed
\begin{align}
P_O &\dot{=} \rho^{-d_o(r)} \text{~~for~~} d_o(r) =\inf_{(v_1,
v_2,u) \in O^+} (v_1+v_2+u) . \label{eq:13}
\end{align}
Comparing \eqref{eq:7} and \eqref{eq:13}, we realize that for
\eqref{eq:4} to be met, $O^{+}$ should be defined as
\begin{align}
O^{+} &= \{(v_1,v_2,u) \in \Real^{3+} |
(\max\{2(1-v_1),1-(v_2+u)\})^+ \leq 2r \} . \nonumber
\end{align}
Then, for any $(v_1,v_2,u) \in O^{c+}$, it is possible to choose $l$
to make $d_e(r,v_1,v_2,u)$ arbitrarily large, ensuring \eqref{eq:4}.
Note that, because of \eqref{eq:9}, $d_o(r)$ provides a lower-bound
on the diversity gain achieved by the protocol. But $d_o(r)$, as
given by \eqref{eq:13}, turns out to be identical to right hand side
of \eqref{eq:36} (refer to \figref{theorem2}). Thus the optimal
diversity-multiplexing tradeoff for this scenario is indeed given by
\eqref{eq:55} and the NAF protocol achieves it.

\subsection{Proof of Theorem~\ref{thrm:4}}
Instead of considering specific codes, in the following we upper
bound the average probability of error over random Gaussian ensemble
of code-books (employed by both the source and relay). Therefore,
averaging is invoked with respect to the fading channel distribution
and the random code-books. It is then straightforward to see that
there is at least one code-book in this ensemble whose average
performance, now with respect only to the fading channel
distribution, is better than the predictions of our upper bounds.
For the single relay DDF protocol, the error probability of the ML
decoder, averaged over the ensemble of Gaussian code-books and
conditioned on a certain channel realization, can be upper bounded
using Bayes' rule to give
\begin{align}
P_{E|g_1,g_2,h} &= P_{E,E_r^c|g_1,g_2,h} + P_{E,E_r|g_1,g_2,h}
\nonumber \\
P_{E|g_1,g_2,h} &\leq P_{E|E_r^c,g_1,g_2,h} + P_{E_r|g_1,g_2,h},
\nonumber
\end{align}
where $E_r$ and $E_r^c$ denote the events that the relay decodes
source's message erroneously and its complement, respectively. The
first step in the proof follows from the channel coding theorem
\cite{CE:79} by observing that if \eqref{eq:61} is met, i.e., if the
mutual information between the signal transmitted by the source and
the signal received by the relay exceeds $lR$, then
$P_{E_r|g_1,g_2,h}$ can be made arbitrarily small, provided that the
code-length is sufficiently large. This means that for any $\epsilon
> 0$ and for a sufficiently large code-length,
\begin{align}
P_{E|g_1,g_2,h} &< P_{E|E_r^c,g_1,g_2,h} + \epsilon. \nonumber
\end{align}
Taking the average over the ensemble of channel realizations gives
\begin{align}
P_{E} &< P_{E|E_r^c} + \epsilon, \nonumber \\
P_{E} &\dot{\leq} P_{E|E_r^c}. \nonumber
\end{align}
This means that the exponential order of $P_{E|E_r^c}$, i.e.,
destination's ML error probability assuming \emph{error-free}
decoding at the relay, provides a lower-bound on the diversity gain
achieved by the protocol. Therefore, we only need to characterize
$P_{E|E_r^c}$, which for the sake of notational simplicity, we will
denote by $P_{E}$ in the sequel. To characterize $P_{E}$, we note
that the corresponding PEP [recalling \eqref{eq:49}] is given by
\begin{align}
P_{PE|g_1,g_2,h} &\leq \left(1+|g_1|^2
\frac{E}{2\sigma_v^2}\right)^{-l'}
\left(1+\left(|g_1|^2+|g_2|^2\right)
\frac{E}{2\sigma_v^2}\right)^{-(l-l')} .\nonumber
\end{align}
Defining $v_1$, $v_2$, and $u$ as the exponential orders of
$1/|g_1|^2$, $1/|g_2|^2$, and $1/|h|^2$, respectively, gives
\begin{align}
P_{PE|v_1,v_2,u} &\dot{\leq}
\rho^{-l[f(1-v_1)^++(1-f)(1-\min\{v_1,v_2\})^+]} \text{~~for~~}
(v_1,v_2,u) \in \Real^{3+} , \nonumber
\end{align}
where $f \triangleq l'/l$. At a rate of $R=r\log{\rho}$ BPCU and a
codeword length of $l$, there are a total of $\rho^{rl}$ codewords.
Thus,
\begin{align}
P_{E,O^c} &\dot{\leq} \rho^{-d_e(r)} \nonumber
\end{align}
for
\begin{align}
d_e(r) &=\inf_{(v_1,v_2,u) \in O^{c+}}
l[f(1-v_1)^++(1-f)(1-\min\{v_1,v_2\})^+-r]+(v_1+v_2+u) \label{eq:11}
\end{align}
Examining \eqref{eq:11}, we realize that for \eqref{eq:4} to hold,
$O^{+}$ should be defined as
\begin{align}
O^{+} &= \{(v_1,v_2,u) \in \Real^{3+} |
f(1-v_1)^++(1-f)(1-\min\{v_1,v_2\})^+ \leq r \} \label{eq:112}
\end{align}
so that it is possible to choose $l$ to make $d_e(r)$ arbitrarily
large, ensuring \eqref{eq:4}. As before, $P_O(R)$ is given by
\eqref{eq:13}, which turns out to be identical to $d(r)$ given by
\eqref{eq:10}. To see this, one needs to consider four different
categories of channel realizations. The first category is when both,
$v_1$ and $v_2$ are greater than one. For this category,
\begin{align}
\inf_{\begin{subarray}{c} (v_1,v_2,u) \in O^+,\\ v_1>1,v_2>1
\end{subarray}} (v_1+v_2+u) &=2. \label{eq:110}
\end{align}
The second category is when $1 \geq v_1 \geq 0$ and $v_2>1$. It is
easy to see from \eqref{eq:112} that for this category,
\begin{align}
\inf_{\begin{subarray}{c} (v_1,v_2,u) \in O^+,\\ 1 \geq v_1 \geq 0,
v_2>1 \end{subarray}} (v_1+v_2+u) &=2-r. \label{eq:111}
\end{align}
The third category to be considered is when $v_1 >1$ and $1 \geq v_2
\geq 0$. Before proceeding further, note that from \eqref{eq:61},
one can show that
\begin{align}
u &= 1 - \frac{r}{f}. \label{eq:63}
\end{align}
This implies that $f \geq r$, since $u$ is nonnegative. Returning
back to \eqref{eq:112}, it is easy to verify that for this category
\begin{align}
v_2 &\geq 1-\frac{r}{1-f}. \label{eq:113}
\end{align}
Now, if $f \geq \max\{r,1-r\}$, then from \eqref{eq:113} and
\eqref{eq:63} we get
\begin{align}
\inf_{\begin{subarray}{c} (v_1,v_2,u) \in O^+,\\ v_1 >1, 1 \geq v_2
\geq 0,\\ f \geq \max\{r,1-r\} \end{subarray}} (v_1+v_2+u) &=\inf_{f
\geq \max\{r,1-r\}} 2-\frac{r}{f}, \nonumber
\end{align}
or
\begin{align}
\inf_{\begin{subarray}{c} (v_1,v_2,u) \in O^+, \\ v_1 > 1, 1 \geq
v_2 \geq 0,
\\ f \geq \max\{r,1-r\} \end{subarray} } (v_1+v_2+u) &= \begin{cases} 1 +
\frac{1-2r}{1-r}, & \frac{1}{2} \geq r \geq 0, \\
1, & 1 \geq r \geq \frac{1}{2}.\end{cases} \label{eq:114}
\end{align}
On the other hand, if $1-r > f \geq r$, then
\begin{align}
\inf_{\begin{subarray}{c} (v_1,v_2,u) \in O^+,\\ v_1 >1, 1 \geq v_2
\geq 0, \\1-r > f \geq r \end{subarray}} (v_1+v_2+u) &=\inf_{1-r > f
\geq r} 3-\frac{r}{1-f}-\frac{r}{f}, \nonumber
\end{align}
\textbf{or}
\begin{align}
\inf_{\begin{subarray}{c} (v_1,v_2,u) \in O^+,\\ v_1 >1, 1 \geq v_2
\geq 0, \\1-r > f \geq r \end{subarray}} (v_1+v_2+u) &=1 +
\frac{1-2r}{1-r} \text{~~for~~} \frac{1}{2} > r \geq 0. \nonumber
\end{align}
This means that $\inf_{O^+} (v_1+v_2+u)$, for the third category, is
indeed given by \eqref{eq:114}. It is noteworthy that the tradeoff
curves given by \eqref{eq:110}, \eqref{eq:111} and \eqref{eq:114},
are all better than the genie-aided tradeoff. In other words, the
diversity gain achieved by this protocol is determined by the fourth
category, where, both $v_1$ and $v_2$ are less than or equal to one.
For this category, one needs to consider two cases (Note that
\eqref{eq:63} is still valid, implying $f \geq r$). The first case,
when $0.5 \geq f \geq r$, is very easy. Referring to
\figref{theorem4a} reveals that, in this case, $\inf_{(v_1,v_2) \in
O^+} v_1+v_2$ and therefore $d_o(r)$ is equal to $2(1-r)$ (the genie
aided tradeoff). The second case, when $f > \max\{r,0.5\}$, is a
little bit more difficult. As can be seen from \figref{theorem4b},
in this case
\begin{align}
\inf_{(v_1,v_2) \in O^+} &= \frac{1-r}{f}. \label{eq:62}
\end{align}
%On the other hand, from \eqref{eq:61}, one can show that
%\begin{align}
%u &= 1 - \frac{r}{f}, \label{eq:63}
%\end{align}
%which implies that $f > \max\{r,0.5\}$ ($u$ is nonnegative and we
%had assumed $f > 0.5$).
From \eqref{eq:63} and \eqref{eq:62}, we
conclude that
\begin{align}
d_o(r) &= \inf_{f > \max\{r,0.5\}} 1 + \frac{1-2r}{f},
\end{align}
which gives \eqref{eq:10}. Again, according to \eqref{eq:9},
$d_o(r)$ provides a lower-bound on the diversity gain achieved by
the protocol. On the other hand, $d_o(r)$ is also an upper bound on
the achieved diversity since: 1) for $0\leq r\leq 0.5$ $d_o(r)$ is
the genie-aided diversity and 2) for $0.5\leq r\leq 1$ it is easy to
see that $v_1=\frac{1-r}{r}+\epsilon$, $v_2=0$ and $u=0$ correspond
to a \emph{channel} outage for any $\epsilon>0$. Thus \eqref{eq:10}
is the diversity achieved by the DDF protocol and the proof is
complete.

\subsection{Proof of Theorem~\ref{thrm:5}}
Inspired by the single-relay case, we use ensembles of Gaussian
code-books at the source and all the relays. To characterize the
diversity-multiplexing tradeoff achieved by the DDF protocol with
$N-1$ relays,  we first label the nodes according to the order in
which they start transmission. That is, the source is labelled as
node $1$, the first relay that starts transmission as node $2$, and
so on. We then use Bayes' rule to upper bound the error probability
of the ML decoder, averaged over the ensemble of Gaussian code-books
and conditioned on a certain channel realization, to get
\begin{align}
P_{E|g_j,h_{ji}} &\leq P_{E|\{E_p^c\}_{p=2}^N,g_j,h_{ji}} +
\sum_{n=2}^N P_{E_n|\{E_p^c\}_{p<n},g_j,h_{ji}},\label{eq:100}
\end{align}
where $E_n, n \in \{2,\cdots,N\}$ denotes the event that node $n$
decodes the source message in error, while $E_n^c$ denotes its
complement. Let us now examine $P_{E_n|\{E_p^c\}_{p<n},g_j,h_{ji}}$,
i.e., the probability that node $n \in \{2,\cdots,N\}$ makes an
error in decoding the source message, assuming error-free decoding
at all previous nodes. It follows from the channel coding theorem
\cite{CE:79}, that if the mutual information between the signals
transmitted by the source and active relays and the signal received
by node $n$ exceeds $lR$, then $P_{E_n|\{E_p^c\}_{p<n},g_j,h_{ji}}$
can be made arbitrarily small, provided that the code-length is
sufficiently large. This means that for any $\epsilon > 0$ and for
sufficiently large code-lengths,
\begin{align}
P_{E_n|\{E_p^c\}_{p<n},g_j,h_{ji}} &< \epsilon, \text{~~} n \in
\{2,\cdots,N\}. \label{eq:101}
\end{align}
Using \eqref{eq:101}, \eqref{eq:100} can be written as
\begin{align}
P_{E|g_j,h_{ji}} &\leq P_{E|\{E_p^c\}_{p=2}^N,g_j,h_{ji}} +
(N-1)\epsilon .\nonumber
\end{align}
Taking the average over the ensemble of channel realizations gives
\begin{align}
P_{E} &< P_{E|\{E_p^c\}_{p=2}^N} + (N-1)\epsilon, \nonumber \\
P_{E} &\dot{\leq} P_{E|\{E_p^c\}_{p=2}^N}. \nonumber
\end{align}
This means that the exponential order of $P_{E|\{E_p^c\}_{p=2}^N}$,
i.e., destination's ML error probability assuming \emph{error-free}
decoding at all of the relays, provides a lower-bound on the
diversity gain achieved by the protocol. Therefore, we only need to
characterize $P_{E|\{E_p^c\}_{p=2}^N}$, which for the sake of
notational simplicity, we will denote by $P_{E}$ in the sequel. To
characterize $P_{E}$, we note that the corresponding PEP, is
upper-bounded by
\begin{align}
P_{PE|g_j,h_{ji}} &\leq \prod_{j=1}^{N} \left[ 1 +
\left(\sum_{i=1}^{j}|g_j|^2\right) \frac{E}{2\sigma_v^2}
\right]^{-l_j}. \nonumber
\end{align}
As before, the gain of the channel that connects the $j^{th}$ node
to the destination is denoted by $g_j$, while the gain of the
channel that connects nodes $i$ and $j$ is denoted by $h_{ji}$. We
use $l_j$ to denote the number of symbol-intervals in the codeword
during which a total of $j$ nodes are transmitting, so that
$\sum_{j=1}^{N} l_j = l$, with $l$ denoting the total codeword
length. Note that $\sum_{j=1}^p l_j$ is the number of
symbol-intervals that relay $p+1$ has to wait, before the mutual
information between its received signal and the signals that the
source and other relays transmit exceeds $lR$. Thus
\begin{align}
\sum_{j=1}^p l_j &\leq \ \min \{ l, \lceil \frac {lR} {\log
(1+|h_{p+1,1}|^2 c \rho)} \rceil \}, \text{~~for~~} N-1 \geq p \geq
1. \label{eq:69}
\end{align}
Defining $v_j$ and $u_{ji}$ as the exponential orders of $g_j$ and
$h_{ji}$, respectively, we have
\begin{align}
P_{PE|v_j,u_{ji}} &\dot{\leq} \rho^{-\sum_{j=1}^{N} l_j
(1-\min\{v_1,\dots,v_j\})^+} . \nonumber
\end{align}
Choosing $R = r\log(\rho)$ for a total of $\rho^{rl}$ codewords,
the following expression for the conditional error probability can
be derived.
\begin{align}
P_{E|v_j,u_{ji}} &\dot{\leq} \rho^{-l\left[\sum_{j=1}^{N}
\frac{l_j}{l}(1-\min\{v_1,\dots,v_j\})^+ - r\right]} . \nonumber
\end{align}
Thus, $O^{+}$ is the set of channel realizations that satisfy
\begin{align}
\sum_{j=1}^{N} \frac{l_j}{l} (1 - \min\{v_1 , \dots, v_j\})^+
&\leq r, \nonumber
\end{align}
which can be simplified to
\begin{align}
1-r &\leq \sum_{j=1}^{N} \frac{l_j}{l} \min\{1, v_1 , \dots, v_j\} .
\label{eq:39}
\end{align}
As before, $P_O(R)$ is characterized by
\begin{align}
P_O(R) &\dot{=} \rho^{-d_o(r)} \text{~~for~~} d_o(r) =
\inf_{O^+}\sum_{j=1}^{N} \left(v_j + \sum_{i < j} u_{ji}\right) .
\label{eq:60}
\end{align}
Defining $\tilde{v}_j \triangleq \min\{v_1, \cdots , v_j\}, j=1,
\cdots , N$ lets us simplify \eqref{eq:39} and \eqref{eq:60} to
\begin{align}
1-r &\leq \sum_{j=1}^{N} \frac{l_j}{l} \min\{1, \tilde{v_j}\}
\label{eq:64}\\
d_o(r) &\geq \inf_{O^+}\sum_{j=1}^{N} \left(\tilde{v}_j + \sum_{i <
j} u_{ji}\right) . \label{eq:65}
\end{align}
From the definition of $\tilde{v_j}$, it follows that
\begin{align}
\tilde{v_1} \geq \tilde{v_2} \geq \cdots \geq \tilde{v_N} \geq 0
\nonumber.
\end{align}
Note that \eqref{eq:69} can also be simplified to
\begin{align}
\sum_{j=1}^p \frac{l_j}{l} &\leq \min\{ 1, \frac{r}{(1-u_{p+1,1})^+}
\}, \text{~~for~~} N-1 \geq p \geq 1, \nonumber
\end{align}
or
\begin{align}
1-\frac{r}{\sum_{k=1}^p \frac{l_k}{l}} &\leq u_{j1}, \quad
\text{~~for~~} j > p. \label{eq:71}
\end{align}

In order to characterize $d_o(r)$, we need to consider three cases.
The first case is when $1 \geq \tilde{v}_1$. In this case,
\eqref{eq:64} simplifies to
\begin{align}
1-r &\leq \sum_{j=1}^{N} \frac{l_j}{l} \tilde{v_j} \nonumber.
\end{align}
Let us define $x_j \triangleq j(\tilde{v}_j - \tilde{v}_{j+1}), j =
1, \cdots, N-1 $ and $x_N \triangleq N \tilde{v}_N$. It immediately
follows that $x_j \geq 0, j=1, \dots, N$. It is also easy to verify
that
\begin{align}
\sum_{j=1}^N \tilde{v}_j &= \sum_{j=1}^N x_j \text{~~and~~} 1-r \leq
\sum_{j=1}^{N} \frac{f_j}{j} x_j, \label{eq:66}
\end{align}
where $f_j \triangleq \sum_{k=1}^{j} l_k/l$. From \eqref{eq:66}, it
can be seen that
\begin{align}
\inf_{\begin{subarray}{c} O^+ \\ 1 \geq \tilde{v}_1 \end{subarray} }
\sum_{j=1}^N \tilde{v}_j &= p(\frac{1-r}{f_p}), \text{~~where~~} p =
\arg \max_{N \geq j \geq 1} \{ \frac{f_j}{j}\} \label{eq:67}.
\end{align}
The infimum value corresponds to $x_p=p(1-r)/f_p$ and $x_j = 0, j
\neq p$ or $\tilde{v}_j = (1-r)/f_p, p \geq j \geq 1$ and
$\tilde{v}_j=0, j
> p$. But we assumed $1 \geq \tilde{v}_1$, so
\begin{align}
f_p &\geq 1-r. \label{eq:68}
\end{align}
From \eqref{eq:71}, it follows that,
\begin{align}
\inf_{\begin{subarray}{c} O^+ \\ 1 \geq \tilde{v}_1 \end{subarray} }
\sum_{j > p} u_{j1} &= (N-p)(1- \frac{r}{f_p}), \quad 1 \geq f_p
\geq r. \label{eq:72}
\end{align}
Now, from \eqref{eq:67} and \eqref{eq:72} we conclude that
\begin{align}
\inf_{\begin{subarray}{c} O^+, \\ 1 \geq \tilde{v}_1
\end{subarray} } \sum_{j=1}^{N} \left(\tilde{v}_j + \sum_{i < j}
u_{ji}\right) &\geq \inf_{\begin{subarray}{c} N \geq p \geq 1, \\
1 \geq f_p \geq max\{r,1-r\} \end{subarray} } d_o(r,p,f_p),
\label{eq:73}
\end{align}
where,
\begin{align}
d_o(r,p,f_p) &\triangleq p(\frac{1-r}{f_p}) + (N-p)(1 -
\frac{r}{f_p}). \label{eq:75}
\end{align}
It turns out that, \eqref{eq:75} is an increasing function of $p$.
Therefore, its infimum corresponds to $p=1$. Now, examining
$d_o(r,1,f_1)$, i.e.,
\begin{align}
d_o(r,1,f_1) &= (\frac{1-r}{f_1}) + (N-1)(1 - \frac{r}{f_1}),
\nonumber
\end{align}
we realize that, for $1/N \geq r \geq 0$, it decreases with $f_1$,
thus its infimum corresponds to $f_1=1$. On the other hand, for $1
\geq r \geq 1/N$, $d_o(r,1,f_1)$ becomes an increasing function of
$f_1$, which means that its infimum corresponds to
$f_1=\max\{r,1-r\}$, i.e.
\begin{align}
\inf_{\begin{subarray}{c} O^+, \\ 1 \geq \tilde{v}_1
\end{subarray} } \sum_{j=1}^{N} \left(\tilde{v}_j + \sum_{i < j}
u_{ji}\right) &\geq
\begin{cases}
N(1-r), & \frac{1}{N} \geq r \geq 0, \\
1+\frac{(N-1)(1-2r)}{1-r}, & \frac{1}{2} \geq r \geq \frac{1}{N}, \\
\frac{1-r}{r}, & 1 \geq r \geq \frac{1}{2}.
\end{cases} \label{eq:76}
\end{align}
The second case to be considered is when $\tilde{v}_i > 1 \geq
\tilde{v}_{i+1}$, $N-1 \geq i \geq 1$. It immediately follows that
\begin{align}
\inf_{\begin{subarray}{c} O^+ \\ \tilde{v}_i > 1 \geq
\tilde{v}_{i+1} \end{subarray}} \sum_{j=1}^i \tilde{v}_j &= i.
\label{eq:77}
\end{align}
In this case, \eqref{eq:64} can be written as
\begin{align}
1-r-f_i \leq \sum_{j=i+1}^N \frac{l_j}{l} \tilde{v}_j \label{eq:78}.
\end{align}
If $f_i \geq 1-r$, then from \eqref{eq:78}, we get
\begin{align}
\inf_{ \begin{subarray}{c} O^+ \\ \tilde{v}_i > 1 \geq
\tilde{v}_{i+1}, \\ f_i \geq \max\{r,1-r\}
\end{subarray}} \sum_{j=i+1}^N \tilde{v}_j &= 0 \label{eq:79}.
\end{align}
On the other hand, from \eqref{eq:71}, it follows that,
\begin{align}
\inf_{u_{j1} \geq 1-\frac{r}{f_i}, j>i } \sum_{j=i+1}^N u_{j1} &=
(N-i)(1- \frac{r}{f_i}), \quad 1 \geq f_i \geq r. \label{eq:80}
\end{align}
Now, from \eqref{eq:77}, \eqref{eq:79} and \eqref{eq:80} one can see
that
\begin{align}
\inf_{\begin{subarray}{c} O^+, \\ \tilde{v}_i > 1 \geq
\tilde{v}_{i+1}, \\ f_i \geq \max\{r,1-r\} \end{subarray} }
\sum_{j=1}^{N} \left(\tilde{v}_j + \sum_{i < j} u_{ji}\right) &\geq
\inf_{\begin{subarray}{c} N-1 \geq i \geq 1, \\ 1 \geq f_i \geq
max\{r,1-r\} \end{subarray} } d_o(r,i,f_i), \nonumber
\end{align}
with
\begin{align}
d_o(r,i,f_i) &\triangleq i+(N-i)(1-\frac{r}{f_i}). \nonumber
\end{align}
The infimum of $d_o(r,i,f_i)$ corresponds to $i=1$ and
$f_i=\max\{r,1-r\}$, i.e.,
\begin{align}
\inf_{\begin{subarray}{c} O^+, \\ \tilde{v}_i > 1 \geq
\tilde{v}_{i+1}, \\ f_i \geq \max\{r,1-r\} \end{subarray}}
\sum_{j=1}^{N} \left(\tilde{v}_j + \sum_{i < j} u_{ji}\right) &\geq
\begin{cases} 1 +
\frac{(N-1)(1-2r)}{1-r}, & \frac{1}{2} \geq r \geq 0, \\
1, & 1 \geq r \geq \frac{1}{2}.\end{cases} \label{eq:85}
\end{align}
If $f_i<1-r$, then the problem of finding $\inf \sum_{j=i+1}^N
\tilde{v}_j$ reduces to the first case (i.e., $1 \geq \tilde{v}_1$).
Specifically, $\inf \sum_{j=i+1}^N \tilde{v}_j$ is given by
\eqref{eq:67}, with $N-i$, $f_p-f_i$, $r+f_i$ and $p-i$ substituting
$N$, $f_p$, $r$ and $p$. Thus,
\begin{align}
\inf_{\begin{subarray}{c} O^+, \\ \tilde{v}_i > 1 \geq
\tilde{v}_{i+1}, \\ 1-r > f_i \geq r \end{subarray}} (\sum_{j=i+1}^N
\tilde{v}_j) &= (p-i)(\frac{1-r-f_i}{f_p-f_i}), \quad
\text{~~where~~} p=\arg \max_{N \geq j \geq i+1}
\{\frac{f_j-f_i}{j-i}\}. \label{eq:81}
\end{align}
Note that \eqref{eq:68} still holds. Derivation of $\inf
\sum_{j=1}^N \sum_{i<j} u_{ji}$ follows from \eqref{eq:71},
\begin{align}
\inf_{u_{j1} \geq 1-\frac{r}{f_k}, j>k } \sum_{j=1}^N \sum_{i<j}
u_{ji} &\geq (p-i)(1-\frac{r}{f_i})+(N-p)(1-\frac{r}{f_p}),
\text{~~with~~} f_p > f_i \geq r. \label{eq:82}
\end{align}
From \eqref{eq:77}, \eqref{eq:81} and \eqref{eq:82}, we conclude
that
\begin{align}
\inf_{\begin{subarray}{c} O^+, \\ \tilde{v}_i > 1 \geq
\tilde{v}_{i+1}, \\ 1-r > f_i \geq r \end{subarray} } \sum_{j=1}^{N}
\left(\tilde{v}_j + \sum_{i < j} u_{ji}\right) &\geq
\inf_{\begin{subarray}{c} N \geq p > i \geq 1, \\ 1 \geq f_p \geq
1-r > f_i \geq r
\end{subarray} } d_o(r,i,p,f_i,f_p), \label{eq:87}
\end{align}
where,
\begin{align}
d_o(r,i,p,f_i,f_p) &\triangleq i + (p-i)(\frac{1-r-f_i}{f_p-f_i}) +
(p-i)(1-\frac{r}{f_i}) + (N-p)(1-\frac{r}{f_p}). \label{eq:83}
\end{align}
As can be seen from \eqref{eq:83}, $d_o(r,i,p,f_i,f_p)$ is a linear,
and therefore monotonic, function of $p$. Thus, its infimum
corresponds to either $p=i+1$ or $p=N$. Now if the infimum indeed
corresponds to $p=i+1$, by plugging in $p=i$ into \eqref{eq:83}, we
derive a lower-bound on it. That is,
\begin{align}
\inf_{\begin{subarray}{c} N \geq p > i \geq 1, \\ 1 \geq f_p \geq
1-r > f_i \geq r \end{subarray} } d_o(r,i,p,f_i,f_p) &\geq
\inf_{\begin{subarray}{c} N > i \geq 1 \\ 1 \geq f_p \geq 1-r
\end{subarray}} i+(N-i)(1- \frac{r}{f_p}). \nonumber
\end{align}
or
\begin{align}
\inf_{\begin{subarray}{c} N \geq p > i \geq 1, \\ 1 \geq f_p \geq
1-r > f_i \geq r \end{subarray} } d_o(r,i,p,f_i,f_p) &\geq
1+(N-1)\frac{1-2r}{1-r}, \text{~~for~~} \frac{1}{2} > r \geq 0.
\label{eq:84}
\end{align}
Choosing $p=N$, on the other hand, gives
\begin{align}
d_o(r,i,N,f_i,1) &= i + (N-i)(2-\frac{r}{1-f_i}-\frac{r}{f_i}),
\nonumber
\end{align}
which has an infimum value, corresponding to $i=1$ and $f_i=r$ or
$f_i=1-r$, identical to the right-hand side of \eqref{eq:84}. This
means that
\begin{align}
\inf_{\begin{subarray}{c} N \geq p > i \geq 1, \\
1 \geq f_p \geq 1-r > f_i \geq r \end{subarray} } d_o(r,i,p,f_i,f_p)
&= 1+(N-1)\frac{1-2r}{1-r}, \text{~~for~~} \frac{1}{2} > r \geq 0.
\label{eq:86}
\end{align}
Now, from \eqref{eq:86}, \eqref{eq:87} and \eqref{eq:85}, we
conclude that
\begin{align}
\inf_{\begin{subarray}{c} O^+, \\ \tilde{v}_i > 1 \geq
\tilde{v}_{i+1} \end{subarray} } \sum_{j=1}^{N} \left(\tilde{v}_j
+ \sum_{i < j} u_{ji}\right) &\geq \begin{cases} 1 +
\frac{(N-1)(1-2r)}{1-r}, & \frac{1}{2} \geq r \geq 0, \\
1, & 1 \geq r \geq \frac{1}{2}.\end{cases} \label{eq:88}
\end{align}
The third case (i.e., $\tilde{v}_N > 1$), is trivial
\begin{align}
\inf_{\begin{subarray}{c} O^+, \\ \tilde{v}_N > 1 \end{subarray} }
\sum_{j=1}^{N} \left(\tilde{v}_j + \sum_{i < j} u_{ji}\right)
&\geq N. \label{eq:89}
\end{align}
From \eqref{eq:76}, \eqref{eq:88} and \eqref{eq:89} we conclude that
\eqref{eq:37} provides a lower-bound on the diversity gain achieved
by the protocol. On the other hand, $d_o(r)$ is also an upper bound
on the diversity since: 1) for $1/N \geq r\geq 0$, $d_o(r)$ is the
genie-aided diversity, 2) for $0.5 \geq r \geq 1/N$, it can be shown
that the realization, where $v_1=1+\epsilon$, $\{v_j\}_{j=2}^N=0$,
$\{u_{j1}\}_{j=2}^N=\frac{1-2r}{1-r}$ and $\{u_{ji}\}_{i \neq j}=0$
corresponds to a \emph{channel} outage for any $\epsilon>0$, and 3)
for $1 \geq r\geq 0.5$, realization $v_1=\frac{1-r}{r}+\epsilon$,
$\{v_j\}_{j=2}^N=0$ and $\{u_{ji}\}=0$ also corresponds to a channel
outage for any $\epsilon>0$. Thus \eqref{eq:37} is the diversity
achieved by the $N-1$ relay DDF protocol and the proof is complete.

\subsection{Proof of Theorem~\ref{thrm:6}}
To characterize the diversity-multiplexing tradeoff achieved by the
CB-DDF protocol, we first label the $N$ destinations according to
the order in which they start transmission. That is, the first
destination that starts transmission is denoted as destination $1$,
the next destination as destination $2$, and so on. Note that the
error probability of destination $j$ can be written as
\begin{align}
P_{E_j} &= P_{E_j|S_j^c}P_{S_j^c} + P_{E_j|S_j}P_{S_j},
\label{eq:42}
\end{align}
where $S_j$ denotes the event that destination $j$ decodes the
message and starts re-transmission before the end of the codeword
and $S_j^c$ is its complement. Now, since both $P_{S_j}$ and
$P_{S_j^c}$ are less than one, \eqref{eq:42} gives
\begin{align}
P_{E_j} &\leq P_{E_j|S_j^c} + P_{E_j|S_j} . \label{eq:43}
\end{align}
In order to characterize $P_{E_j|S_j}$, we need to characterize
$P_{E_j|S_j,g,h}$, i.e., destination $j$'s ML error probability,
averaged over the ensemble of Gaussian code-books and conditioned on
a certain channel realization, under the assumption that it started
transmission before the end of the codeword. Towards this end and
through using Bayes' rule, one can upper bound $P_{E_j|S_j,g,h}$ to
get
\begin{align}
P_{E_j|S_j,g,h} &\leq \sum_{i=1}^j
P_{E_i|\{E_p^c\}_{p<i},S_j,g,h}.\label{eq:102}
\end{align}
Now, let us examine $P_{E_i|\{E_p^c\}_{p<i},S_j,g,h}$, i.e., the
probability that destination $i$ ($i\leq j$), makes an error in
decoding the source message, conditioned on $S_j$ (which ensures
that destination $i$ has indeed started re-transmission) and
assuming \emph{error-free} decoding at all of the active
destinations. It follows from the channel coding theorem
\cite{CE:79}, that if the mutual information between the signals
transmitted by the source and active destinations and the signal
received by destination $i$ exceeds $lR$ (which is implied by
$S_j$), then $P_{E_i|\{E_p^c\}_{p<i},S_j,g,h}$ can be made
arbitrarily small, provided that the code-length is sufficiently
large. This means that for any $\epsilon > 0$ and for sufficiently
large code-lengths,
\begin{align}
P_{E_i|\{E_p^c\}_{p<i},S_j,g,h} &< \epsilon, \text{~~} i \leq j.
\label{eq:103}
\end{align}
Using \eqref{eq:103}, \eqref{eq:102} can be written as
\begin{align}
P_{E_j|S_j,g,h} &\leq j\epsilon .\nonumber
\end{align}
Taking the average over the ensemble of channel realizations gives
\begin{align}
P_{E_j|S_j} &< j\epsilon. \nonumber
\end{align}
This together with \eqref{eq:43}, yields
\begin{align}
P_{E_j} &< P_{E_j|S_j^c} + j\epsilon, \nonumber \\
P_{E_j} &\dot{\leq} P_{E_j|S_j^c}. \label{eq:104}
\end{align}
This means that the exponential order of $P_{E_j|S_j^c}$, provides a
lower-bound on the diversity gain achieved by the protocol. Now,
examining $P_{E_j|S_j^c}$, it is easy to realize that the event in
which the $j^{th}$ destination (out of $N$ destinations), spends the
entire codeword listening, i.e. $S_j^c$, is identical to the DDF
relay protocol with the rest of the destinations taking the role of
the $N-1$ relays. Thus, from \eqref{eq:104}, we see that
communication to the $j^{th}$ destination achieves the same
diversity order as does the DDF relay protocol with $N-1$ relays,
namely, \eqref{eq:41}. This completes the proof.

\subsection{Proof of Theorem~\ref{thrm:7}}
Realizing that \eqref{eq:14} also corresponds to the optimal
diversity-multiplexing tradeoff for a MIMO point-to-point
communication system with $N$ transmit antennas and a single receive
antenna (i.e., the case of ``genie-aided'' cooperation between $N$
sources), we only need to show that the CMA-NAF protocol achieves
this tradeoff. To achieve this goal, we assume that each of the
sources uses a Gaussian random code with codeword length $l$ and
data rate $R$, where $l$ is chosen as in \eqref{eq:15} and $R$ grows
with $\rho$ according to \eqref{eq:58}. We then characterize the
joint ML decoder's error probability $P_E(\rho)$. Note that the
error probability of the joint ML decoder upper-bounds the error
probabilities of the source-specific ML decoders and thus provides a
lower-bound on the achievable overall diversity gain (as a function
of $r$). In characterizing $P_E(\rho)$, we follow the approach of
Tse {\em et al.} \cite{TVZ:03} by partitioning the error event $E$
into the set of partial error events $\{E^I\}$, i.e.,
\begin{align}
E &= \bigcup_I E^I , \nonumber
\end{align}
where $I$ denotes any \emph{nonempty} subset of $\{1,...,N\}$ and
$E^I$ (referred to as a ''type-$I$ error'') is the event that the
joint ML decoder incorrectly decodes the messages from sources whose
indices belong to $I$ while correctly decoding all other messages.
Because the partial error events are mutually exclusive,
\begin{align}
P_E(\rho) &= \sum_I P_{E^I}(\rho) . \label{eq:16}
\end{align}

Using Bayes' rule, one can upper-bound $P_{E^I}(\rho)$ as
\begin{align}
P_{E^I}(\rho) &= P_O(R)P_{E^I|O}+P_{E^I,O^c} \nonumber \\
P_{E^I}(\rho) &\leq P_O(R)+P_{E^I,O^c} , \nonumber
\end{align}
where, as before, $O$ and $O^c$ denote the outage event and its
complement, respectively. The outage event is defined such that
$P_O(R)$ dominates $P_{E^I,O^c}$ for all $I$:
\begin{align}
P_{E^I,O^c} &\dot{\leq} P_O(R) . \label{eq:17}
\end{align}
Thus,
\begin{align}
P_{E^I}(\rho) &\dot{\leq} P_O(R), \nonumber
\end{align}
which, together with \eqref{eq:16}, results in
\begin{align}
P_E(\rho) &\dot{\leq} P_O(R) . \label{eq:18}
\end{align}
This means that $P_O(R)$, as defined by \eqref{eq:17}, provides an
upper-bound to the joint ML decoder's error probability and
therefore a lower-bound to the achievable diversity gain $d^*(r)$.
The derivation of $P_O(R)$, however, requires the characterization
of $P_{PE^I|g_j,h_{ji}}$ (i.e., the joint ML decoder's type-$I$ PEP,
conditioned on a particular channel realization and averaged over
the ensemble of Gaussian random codes). Here, we upper-bound
$P_{PE^I|g_j,h_{ji}}$, for each $I$, by the PEP of a suboptimal
joint ML decoder that uses only a subset of the destination's
observations (referred to as the \emph{type-$I$ decoder}):
\begin{align}
P_{PE^I|g_j,h_{ji}} &\leq
\det(I_m+\frac{1}{2}\Sigma_{\mathbf{s}^I}\Sigma_{\mathbf{n}^I}^{-1})^{-1}
\label{eq:20}
\end{align}
In \eqref{eq:20}, $\Sigma_{\mathbf{s}^I}$ and
$\Sigma_{\mathbf{n}^I}$ represent the $m\times m$ covariance
matrices corresponding to the signal and noise components,
respectively, of the \emph{partial} observation vector used by the
type-$I$ decoder, provided that the symbols of the sources that are
not in set $I$ are set to zero. The size $m$ will be characterized
in the sequel.

Before going into more detail on the type-$I$ decoder, we note that,
since $\Sigma_{\mathbf{s}^I}$ and $\Sigma_{\mathbf{n}^I}$ are both
positive definite matrices, the right-hand side of \eqref{eq:20} can
be upper-bounded as
\begin{align}
P_{PE^I|g_j,h_{ji}} &\dot{\leq}
\det(\Sigma_{\mathbf{s}^I})^{-1}\det(\Sigma_{\mathbf{n}^I}) .
\label{eq:21}
\end{align}
The discussion is simplified if we define $v_j$ and $u_{ji}$ as the
exponential orders of $1/|g_j|^2$ and $1/|h_{ji}|^2$, respectively.
Note that the exponential orders of $\{|b_j|^2\}_{j=1}^{N}$ do not
appear in the following expressions for the reasons outlined in the
proof of Theorem~\ref{thrm:2}. We also note that the exponential
orders of the broadcast gains $\{|a_j|^2\}_{j=1}^N$ are zero.
Furthermore, recalling \eqref{eq:47}, the PDFs of negative $v_j$ and
$u_{ji}$ are effectively zero for large values of $\rho$, allowing
us to concern ourselves only with their non-negative realizations.
With this ideas in mind, we return to \eqref{eq:21} and claim that
\begin{align}
\det(\Sigma_{\mathbf{n}^I}) &\dot{\leq} 1 . \label{eq:23}
\end{align}
To understand \eqref{eq:23}, recall that the noise component of the
destination observation is a linear combination of the noise
originating at the sources (i.e., $\{w_{j,k}\}_{j=1}^{N}$) and the
noise originating at the destination (i.e., $v_{j,k}$). Furthermore,
the coefficients of this linear combination are the products of some
channel, broadcast, and repetition gains.  Then, because these noise
variances and magnitude-squared gains can be written as non-positive
powers of $\rho$, equation \eqref{eq:23} must hold. Combining
\eqref{eq:23} and \eqref{eq:21} yields
\begin{align}
P_{PE^I|v_j,u_{ji}} &\dot{\leq} \det( \Sigma_{\mathbf{s}^I} )^{-1}
\text{~~for~~} v_j \geq 0, u_{ji} \geq 0 . \label{eq:24}
\end{align}

As mentioned earlier, $\Sigma_{\mathbf{s}^I}$ represents the
covariance matrix of the signal component of the partial observation
used by the type-$I$ decoder, provided that the symbols of the
sources that are not in $I$ are set to zero. To fully characterize
$\Sigma_{\mathbf{s}^I}$, though, we must know which observations are
used by the type-$I$ decoder and which are discarded. The type-$I$
decoder picks \emph{one} observation for every source in set $I$,
for a total of $m=|I|$ observations per frame (where $|I|$ denotes
the size of $I$ and therefore $1\leq |I|\leq N$). Provided that
frame $k$ is not the last frame in its super-frame and assuming that
during this super-frame, source $i$ is helping source $j\in I$, the
destination observation component corresponding to source $j$ will
be either the $y_{j,k}$ that corresponds to source $j$'s broadcast
of $x_{j,k}$ or the $y_{i,k'}$ that corresponds to helper $i$'s
re-broadcast of $x_{j,k}$ (where $k'\in\{k,k+1\}$). As an example,
consider the case when $N=4$ and assume that during a certain
super-frame, source $3$ is helping source $2 \in I$ (i.e., $j=2$,
$i=3$). In this case, the type-$I$ decoder picks either $y_{2,k}$ or
$y_{3,k}$ in correspondence to $x_{2,k}$. However, if instead of
source $3$, source $1$ is helping source $2$ (i.e., $j=2$, $i=1$),
then the type-$I$ decoder has to choose between $y_{2,k}$ or
$y_{1,k+1}$. Back to our description of the type-$I$ decoder, if $i
\in I$, then the decoder always picks $y_{j,k}$ over $y_{i,k'}$. On
the other hand, if $i \notin I$, then the decoder chooses $y_{j,k}$
when $|g_j|^2 \geq |g_i|^2$ or $y_{i,k'}$ when $|g_j|^2 < |g_i|^2$
(i.e., the observation received through the better channel). The
preceding discussion focused on the case where frame $k$ is not the
last frame of the super-frame. If frame $k$ is indeed last, then the
decoder always chooses $y_{j,k}$ over $y_{i,k'}$.

We define $\mathbf{s}^I_{j,k}$, where $j \in I$, as the vector (of
dimension $ml \times 1$) of contributions of symbol $x_{j,k}$ to the
destination observations picked by the type-$I$ decoder. Clearly,
\begin{align}
\mathbf{s}^I &= \sum_{k=1}^{l} \sum_{j \in I} \mathbf{s}_{j,k}^I.
\nonumber
\end{align}
Taking into account the independence of the transmitted symbols
(i.e., $x_{j,k}$), we have
\begin{align}
\Sigma_{\mathbf{s}^I} &= \sum_{k=1}^l \sum_{j\in I} \text{E}
\{\mathbf{s}^I_{j,k}(\mathbf{s}^I_{j,k})^H\}. \label{eq:59}
\end{align}
In order to illuminate some of the properties of
$\mathbf{s}^I_{j,k}$, assume that we sort the chosen observations in
chronological order. From the description given, it is apparent
that, associated with each chosen observation (i.e., $y_{j,k}$ or
$y_{i,k'}$) there is \emph{one} symbol $x_{j,k}$ (with $j \in I$)
which has contributions only from this observation forward. This
means that if we define $S^I$ as
\begin{align}
S^I &\triangleq [\mathbf{s}^I_{j_1,k_1} \mathbf{s}^I_{j_2,k_2} \dots
\mathbf{s}^I_{j_{ml},k_{ml}}]_{ml \times ml}, \nonumber
\end{align}
where $j_p \in I$ and $k_p \in \{1,\dots,l\}$ are chosen such that
the first non-zero elements of $\mathbf{s}^I_{j_p,k_p},
p=1,\dots,ml$ are sorted in chronological order, then $S^I$ will be
lower-triangular and consequently $(S^I)^H$ will be
upper-triangular. Furthermore, based on the choice between $y_{j,k}$
or $y_{i,k{'}}$ (corresponding to $x_{j,k}$), the first non-zero
element of $\mathbf{s}^I_{j_p,k_p}$ (i.e., the $p^{\text{th}}$
diagonal element of $S^I$) will be $g_ja_jx_{j,k}$ or
$g_ib_ih_{ij}a_jx_{j,k}$, respectively. Next, we define
$\mathbf{\psi}^I_{j,k}$ as the signature of $x_{j,k}$, i.e.,
\begin{align}
\mathbf{\psi}^I_{j,k} &\triangleq \frac{1}{x_{j,k}}
\mathbf{s}^I_{j,k} \quad j \in I, \nonumber
\end{align}
and $\Psi^I$ as
\begin{align}
\Psi^I &\triangleq [\mathbf{\psi}^I_{j_1,k_1}
\mathbf{\psi}^I_{j_2,k_2} \dots \mathbf{\psi}^I_{j_{ml},k_{ml}}]_{ml
\times ml}. \nonumber
\end{align}
It follows then, that $\Psi^I$ is also lower-triangular with the
$p^{\text{th}}$ diagonal element being equal to $g_ja_j$ or
$g_ib_ih_{ij}a_j$. Using these definitions, \eqref{eq:59} can be
written as
\begin{align}
\Sigma_{\mathbf{s}^I} &= E \sum_{k=1}^l \sum_{j\in I}
\mathbf{\psi}^I_{j,k}(\mathbf{\psi}^I_{j,k})^H \label{eq:90}.
\end{align}
The significance of $\Psi^I$ can now be seen from the fact that
\eqref{eq:90} can be written as
\begin{align}
\Sigma_{\mathbf{s}^I} &= E\Psi^I(\Psi^I)^H . \nonumber
\end{align}
Now, as the determinant of triangular matrices is simply the product
of their diagonal elements, from \eqref{eq:24} we conclude that
\begin{align}
P_{PE^I|v_j,u_{ji}} &\dot{\leq} \rho^{-m(N-1)L+\sum_{j \in
I}\big[(m-1)Lv_j+\sum_{i \notin
I}\big(\min\{v_j,u_{ji}+v_i\}(L-1)+v_j\big)\big]}, & v_j & \geq 0,
u_{ji} \geq 0 . \nonumber
\end{align}
It is obvious that for large $L$'s, the previous inequality can be
rewritten as
\begin{align}
P_{PE^I|v_j,u_{ji}} &\dot{\leq} \rho^{-\big[-\sum_{j \in I}
\big((m-1)v_j+\sum_{i \notin I} \min\{v_j,u_{ji}+v_i\}
\big)+m(N-1)\big]L}, & v_j & \geq 0, u_{ji} \geq 0 .
\end{align}
At rate $R=r\log(\rho)$ and codeword length $l$, and when the
symbols of the sources that are not in $I$ are set to zero, there
are a total of $\rho^{m(N-1)Lr}$ unique codewords. Thus,
\begin{align}
P_{E^I|v_j,u_{ji}} &\dot{\leq} \rho^{-\big[-\sum_{j \in
I}\big((m-1)v_j+\sum_{i \notin
I}\min\{v_j,u_{ji}+v_i\}\big)+m(N-1)(1-r)\big]L}, & v_j & \geq 0,
u_{ji} \geq 0 . \label{eq:25}
\end{align}
This conditional type-$I$ error probability leads to
\begin{align}
P_{E^I,O^c} &\dot{\leq} \rho^{-d_{e^I}(r)} , \nonumber
\end{align}
where
\begin{align}
d_{e^I}(r) \triangleq& \min_{O_c^{+}} \sum_j\left(v_j +
\sum_iu_{ji}\right)
+ \cdots \nonumber \\
& \left[ -\sum_{j \in I} \left((m-1)v_j + \sum_{i \notin I}
\min\{v_j,u_{ji}+v_i\} \right) + m(N-1)(1-r) \right]L \label{eq:26}
\end{align}
Examining \eqref{eq:26}, we realize that for \eqref{eq:17} to be
met, $O^{+}$ should be defined as the set of all real
$\frac{N(N+1)}{2}$-tuples with nonnegative elements that satisfy the
following condition for at least one nonempty $I \subseteq
\{1,\dots,N\}$:
\begin{align}
\sum_{j \in I}\left((m-1)v_j+\sum_{i \notin
I}\min\{v_j,v_i+u_{ji}\}\right) &\geq m(N-1)(1-r) \label{eq:91}
\end{align}
This way, by choosing large enough $l$, $d_{e^I}(r)$ can be made
arbitrary large and thus \eqref{eq:17} is always met. From
\eqref{eq:91}, it follows that
\begin{align}
\sum_{j \in I}\left((m-1)v_j+\sum_{i \notin I} \min \{v_j,
v_i+\max_{j \neq i} \{u_{ji}\}\} \right) & \ge m(N-1)(1-r).
\label{eq:92}
\end{align}
Substituting $\min \{v_j, v_i+\max_{j \neq i} \{u_{ji}\}\}$ in this
expression by $v_j$ gives
\begin{align}
\sum_{j \in I} v_{j} &\ge m(1-r) \label{eq:93}.
\end{align}
On the other hand, replacing $\min \{v_j, v_i+\max_{j \neq i}
\{u_{ji}\}\}$ in \eqref{eq:92} by $v_i+\max_{j \neq i} \{u_{ji}\}$
results in
\begin{align}
(m-1) \sum_{j \in I} v_j + m \sum_{i \notin I} (v_i+\max_{j \neq i}
\{u_{ji}\}) &\ge m(N-1)(1-r) \label{eq:94}.
\end{align}
Under the constraints given by \eqref{eq:93} and \eqref{eq:94}, it
is easy to see that
\begin{align}
\inf_{O^+} \left(\sum_{j \in I} v_j + \sum_{i \notin I} (v_i+\max_{j
\neq i} \{u_{ji}\})\right) \geq N(1-r). \label{eq:95}
\end{align}
Now, from \eqref{eq:95} and \eqref{eq:60}, it follows that
\begin{align}
d_o(r) &\geq N(1-r). \nonumber
\end{align}
Again, according to \eqref{eq:18}, $d_o(r)$ provides a lower-bound
on the diversity gain achieved by the protocol. Thus the protocol
achieves the diversity gain given by \eqref{eq:14} and the proof is
complete.

\newpage
\putFrag{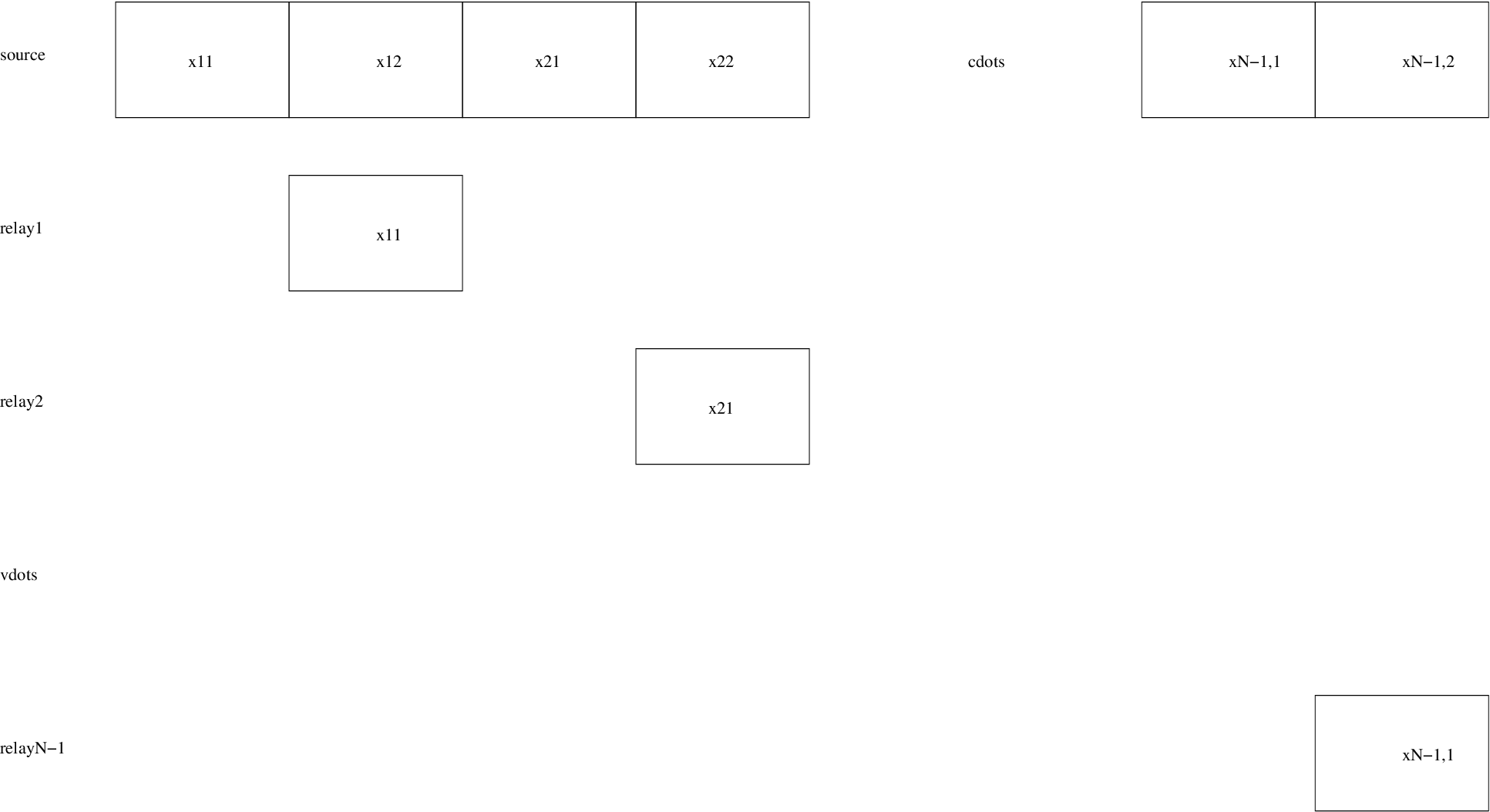}{The super-frame in the NAF protocol with
$N-1$
  relays.}{5}{\footnotesize
\psfrag{source}[]{Source:} \psfrag{x11}[]{$x_{1,1}$}
\psfrag{x12}[]{$x_{2,1}$} \psfrag{x21}[]{$x_{1,2}$}
\psfrag{x22}[]{$x_{2,2}$} \psfrag{cdots}[]{$\cdots$}
\psfrag{xN-1,1}[]{$x_{1,N-1}$} \psfrag{xN-1,2}[]{$x_{2,N-1}$}
\psfrag{relay1}[]{Relay $1$:} \psfrag{relay2}[]{Relay $2$:}
\psfrag{vdots}[]{$\vdots$} \psfrag{relayN-1}[]{Relay $N-1$:} }

\putFrag{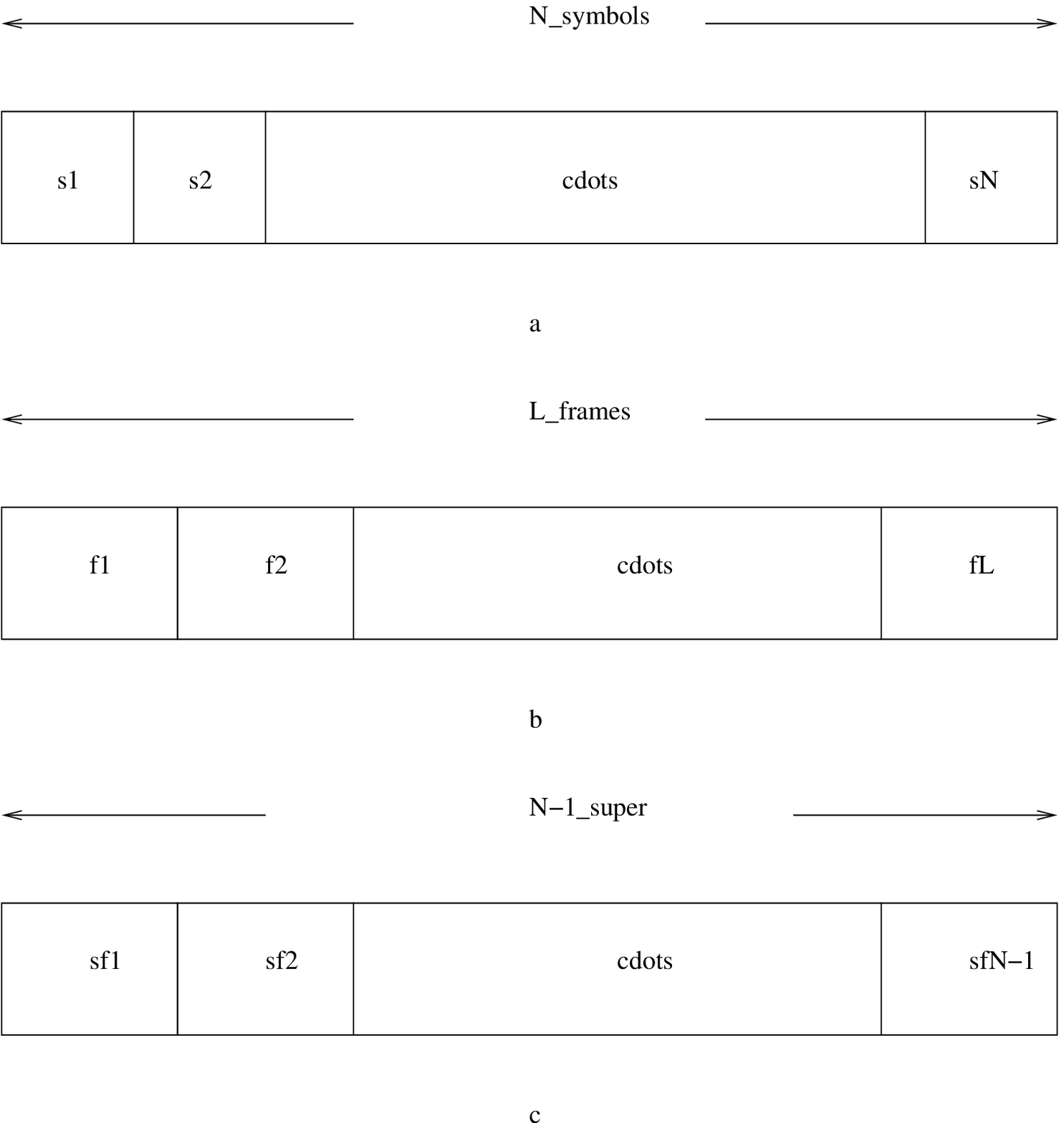}{The cooperation frame, super-frame and
coherence-interval in the CMA-NAF protocol with $N$
  sources.}{3}{\footnotesize
\psfrag{cdots}[]{$\cdots$} \psfrag{a}[]{a) Cooperation Frame.}
\psfrag{N_symbols}[]{$N$ Symbols} \psfrag{s1}[]{$1$}
\psfrag{s2}[]{$2$} \psfrag{sN}[]{$N$} \psfrag{b}[]{b) Super-Frame.}
\psfrag{L_frames}[]{$L$ Frames} \psfrag{f1}[]{$1$}
\psfrag{f2}[]{$2$} \psfrag{fL}[]{$L$} \psfrag{c}[]{c) Coherence
Interval} \psfrag{N-1_super}[]{$N-1$ Super-Frames}
\psfrag{sf1}[]{$1$} \psfrag{sf2}[]{$2$} \psfrag{sfN-1}[]{$N-1$} }

\putFrag{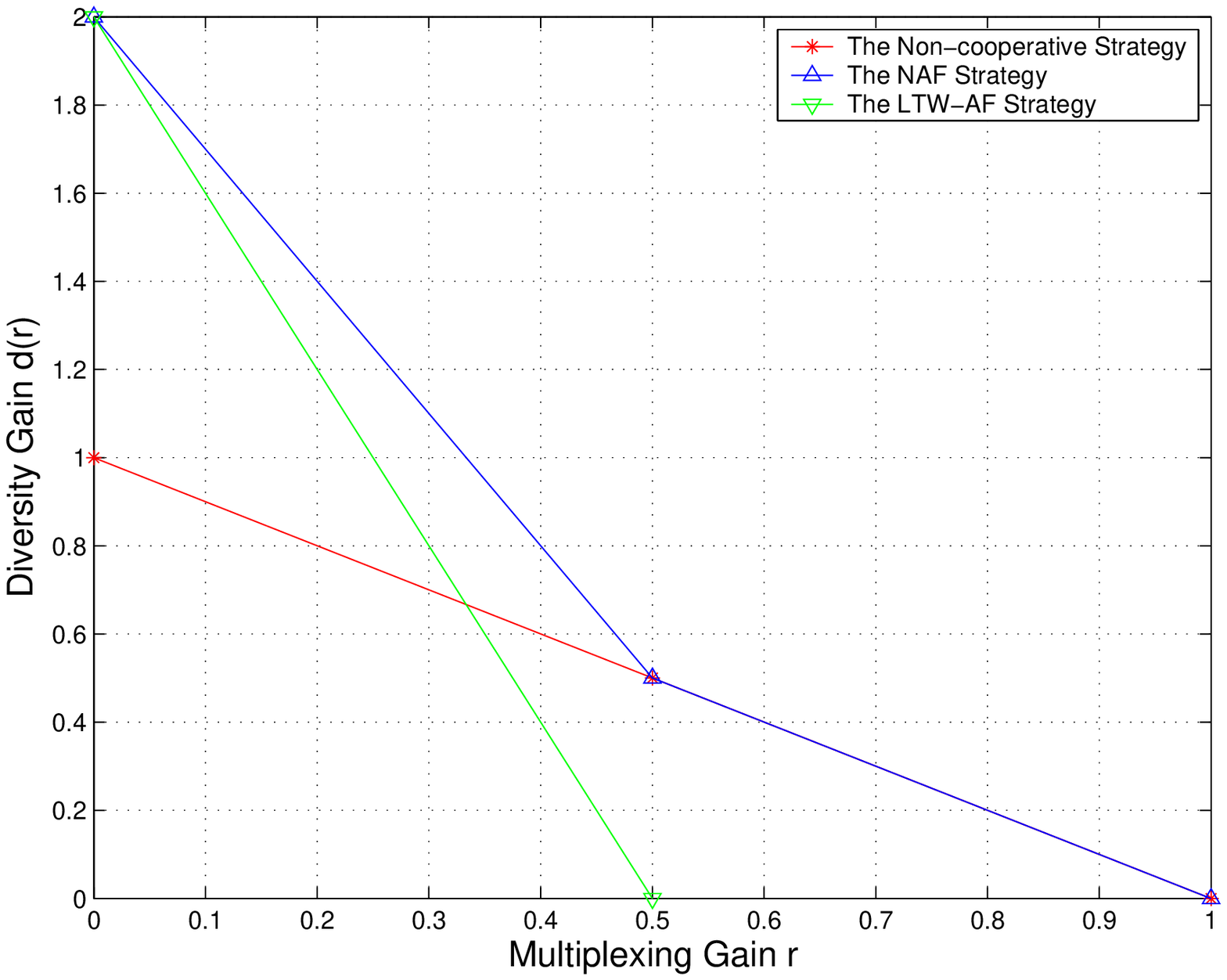}{Optimal diversity-multiplexing tradeoff for a
single-relay AF protocol.}{3}{}

\putFrag{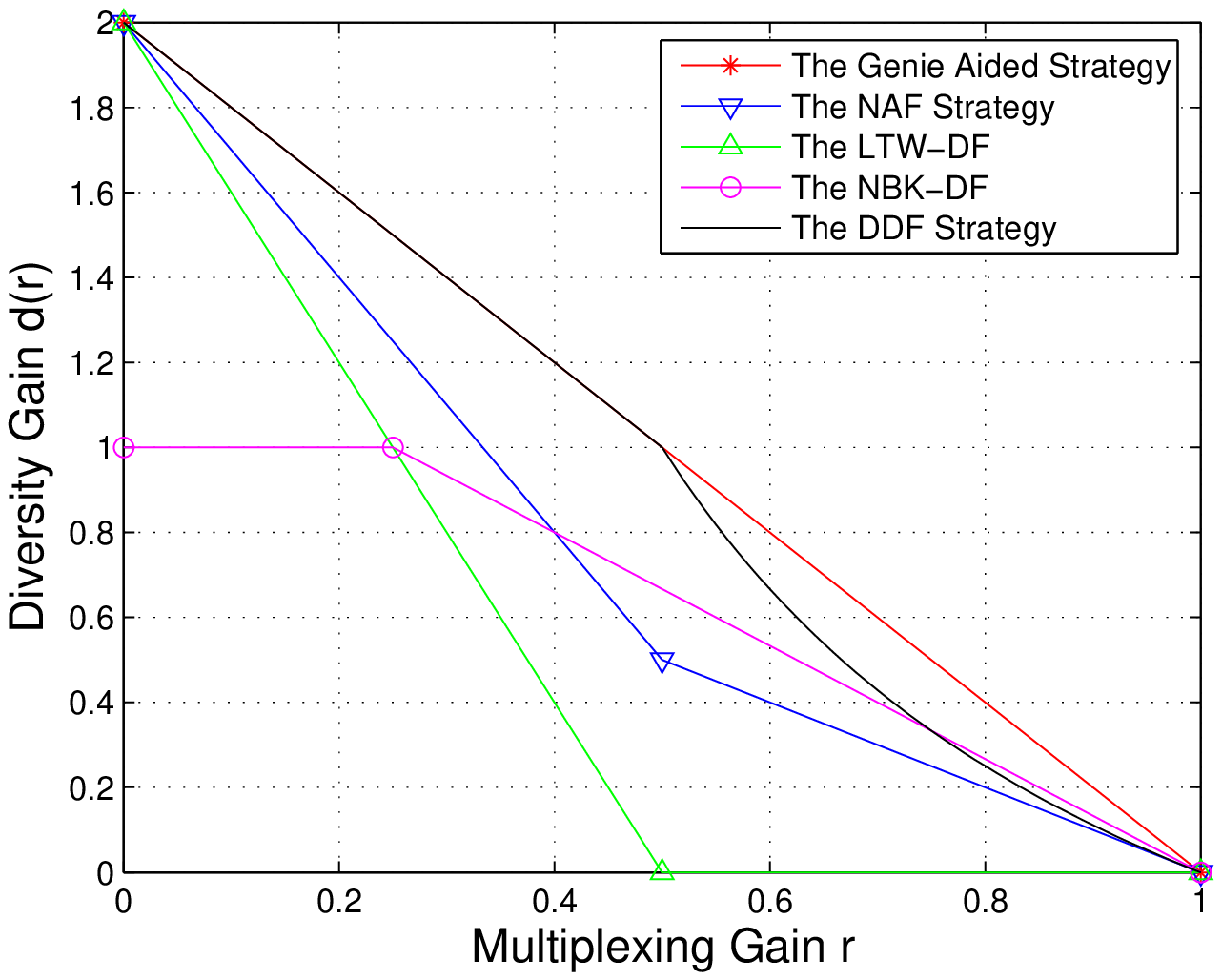}{Diversity-multiplexing tradeoff for the DDF
protocol with one relay.}{3.5}{}

\putFrag{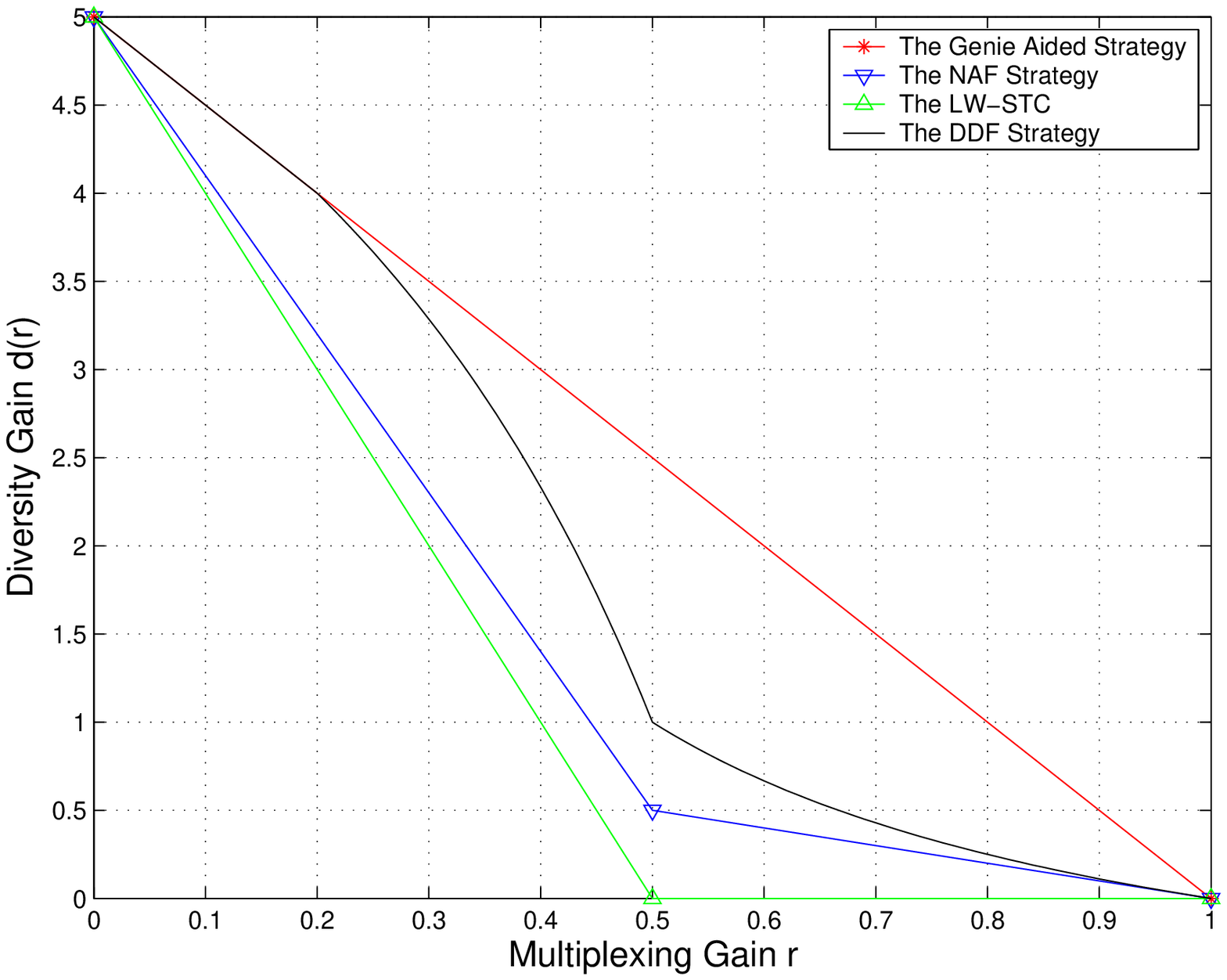}{Diversity-multiplexing tradeoff for the NAF,
DDF, LW-STC, and genie aided protocols with $4$ relays.}{3}{}

\putFrag{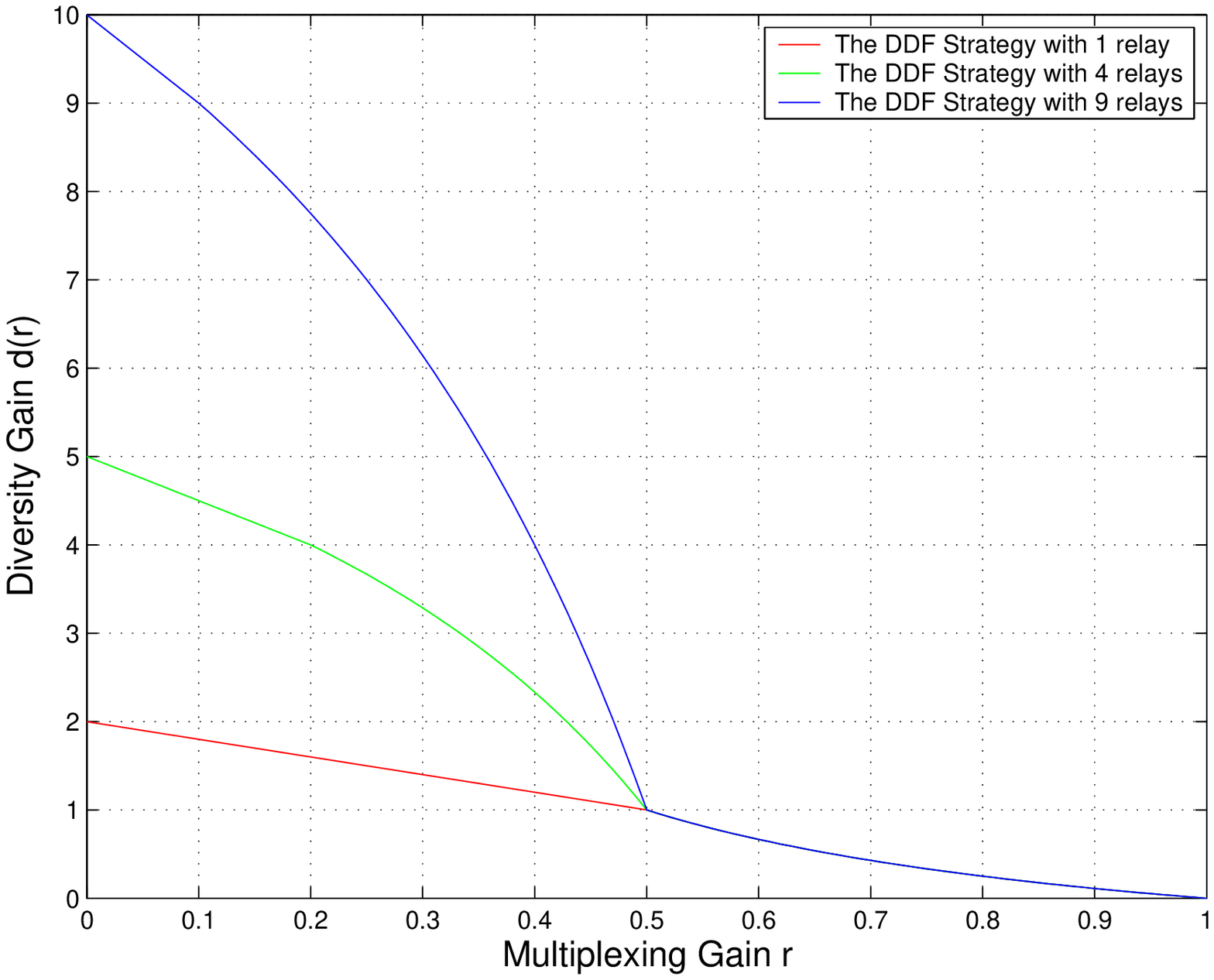}{Diversity-multiplexing tradeoff for the DDF
protocol with different number of relays.}{3}{}

\putFrag{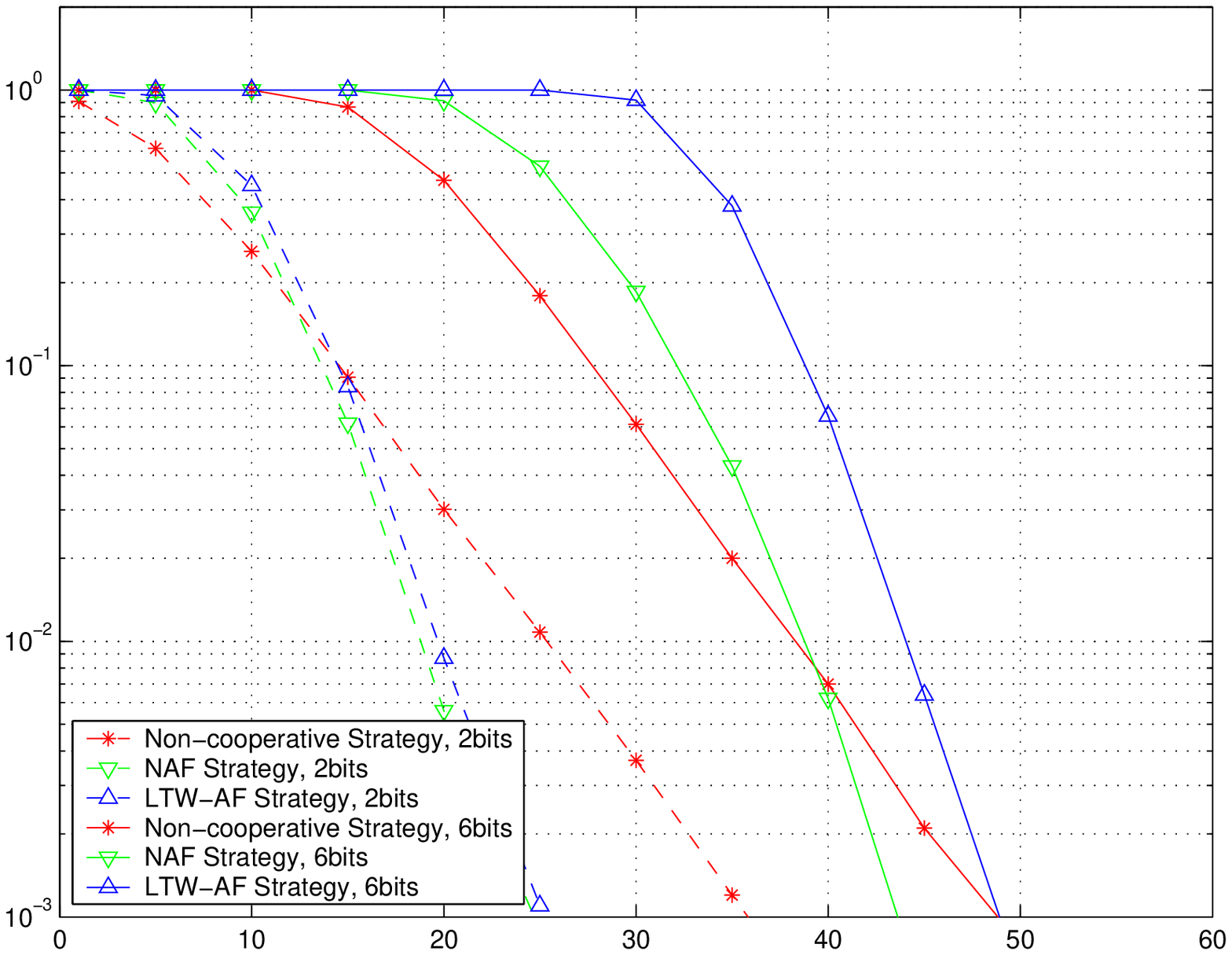}{Comparison of the outage probability for the NAF
relay, LTW-AF, and non-cooperative $1\times1$ protocols
($N=2$).}{3}. {}

\putFrag{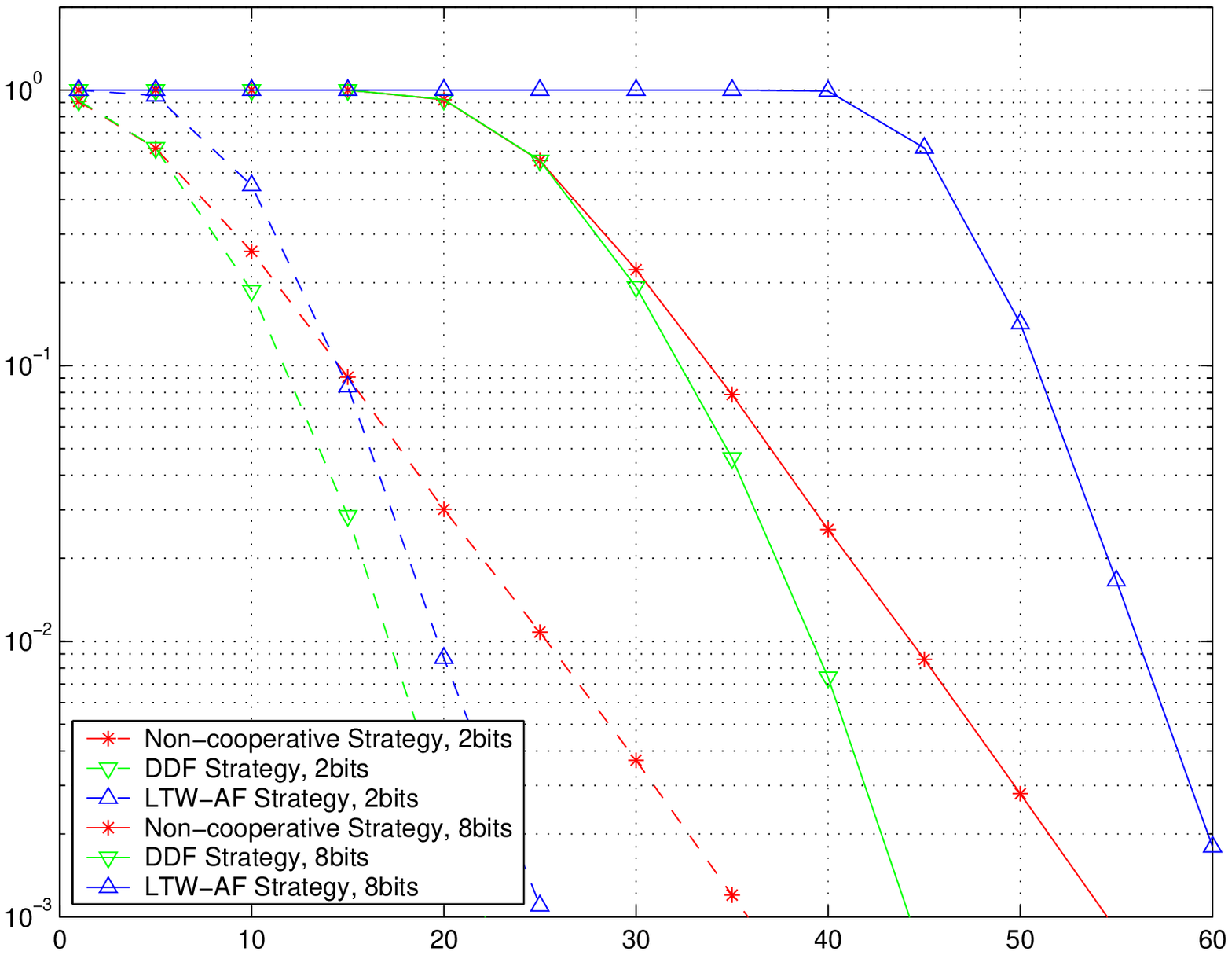}{Comparison of the outage probability for the DDF
relay, LTW-AF and non-cooperative $1\times1$ protocols ($N=2$).}{3}.
{}

\putFrag{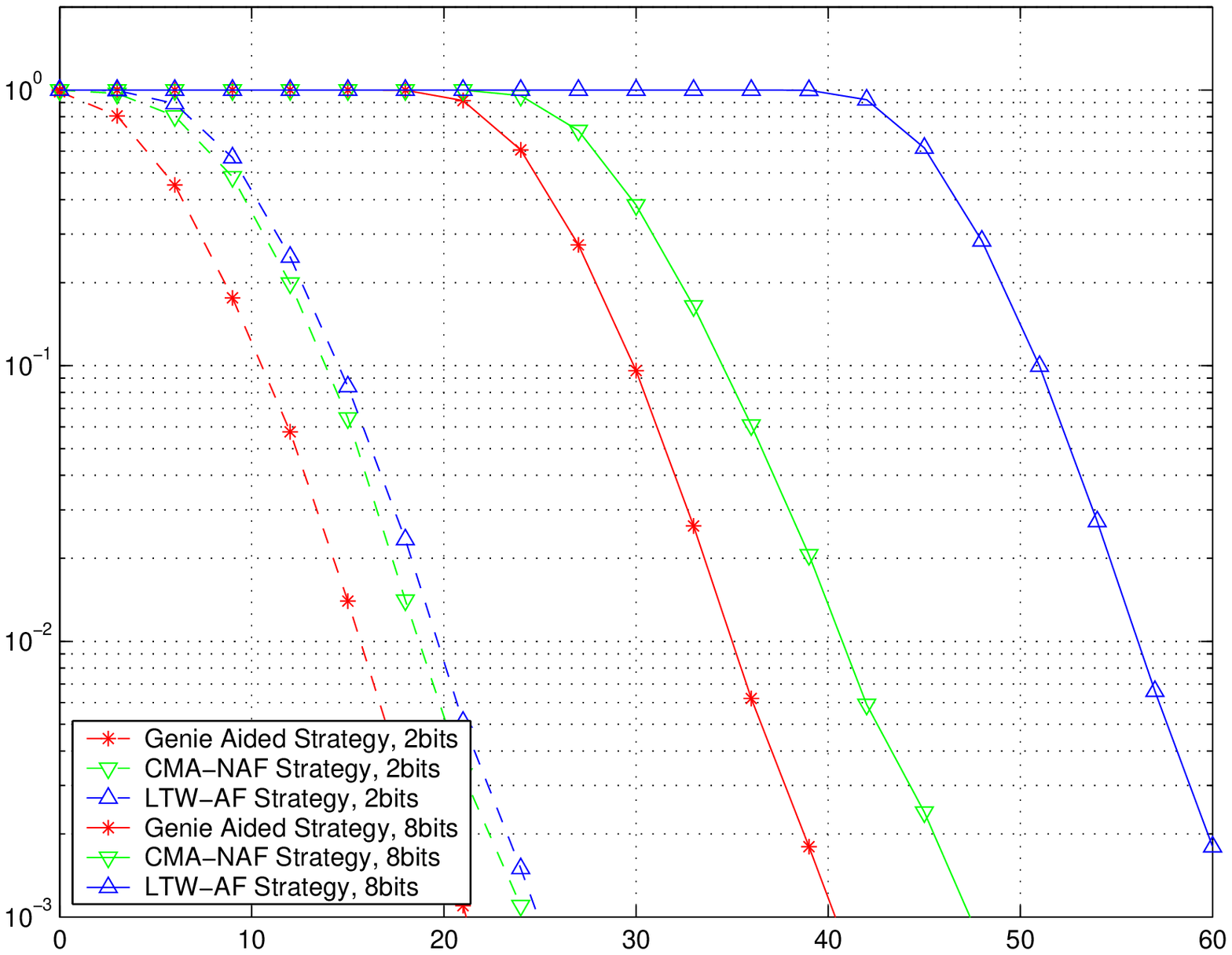}{Comparison of the outage probability for the
CMA-NAF, LTW-AF and genie-aided $2\times1$ protocols ($N=2$).}{3}.
{}

\putFrag{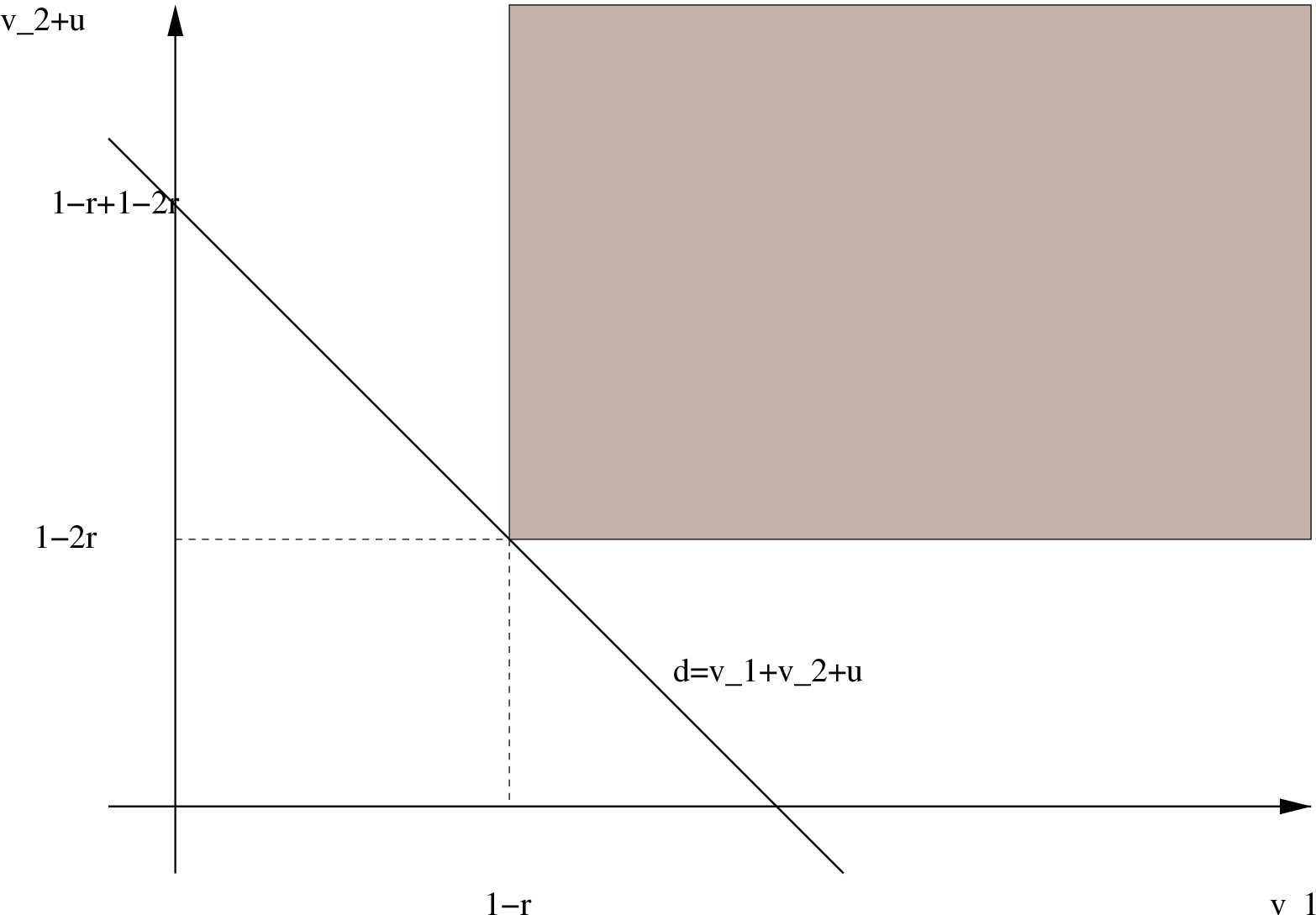}{Outage Region for the NAF protocol with a single
relay.}{3} {\footnotesize
 \psfrag{Outage}[]{$O^+$}
 \psfrag{v_1}[]{$v_1$}
 \psfrag{1-r}[]{$1-r$}
 \psfrag{1-2r}[r]{$(1-2r)^+$}
 \psfrag{1-r+1-2r}[r]{$1-r+(1-2r)^+$}
 \psfrag{d=v_1+v_2+u}[l]{$d=v_1+v_2+u$}
 \psfrag{v_2+u}[r]{$v_2+u$}}

\putFrag{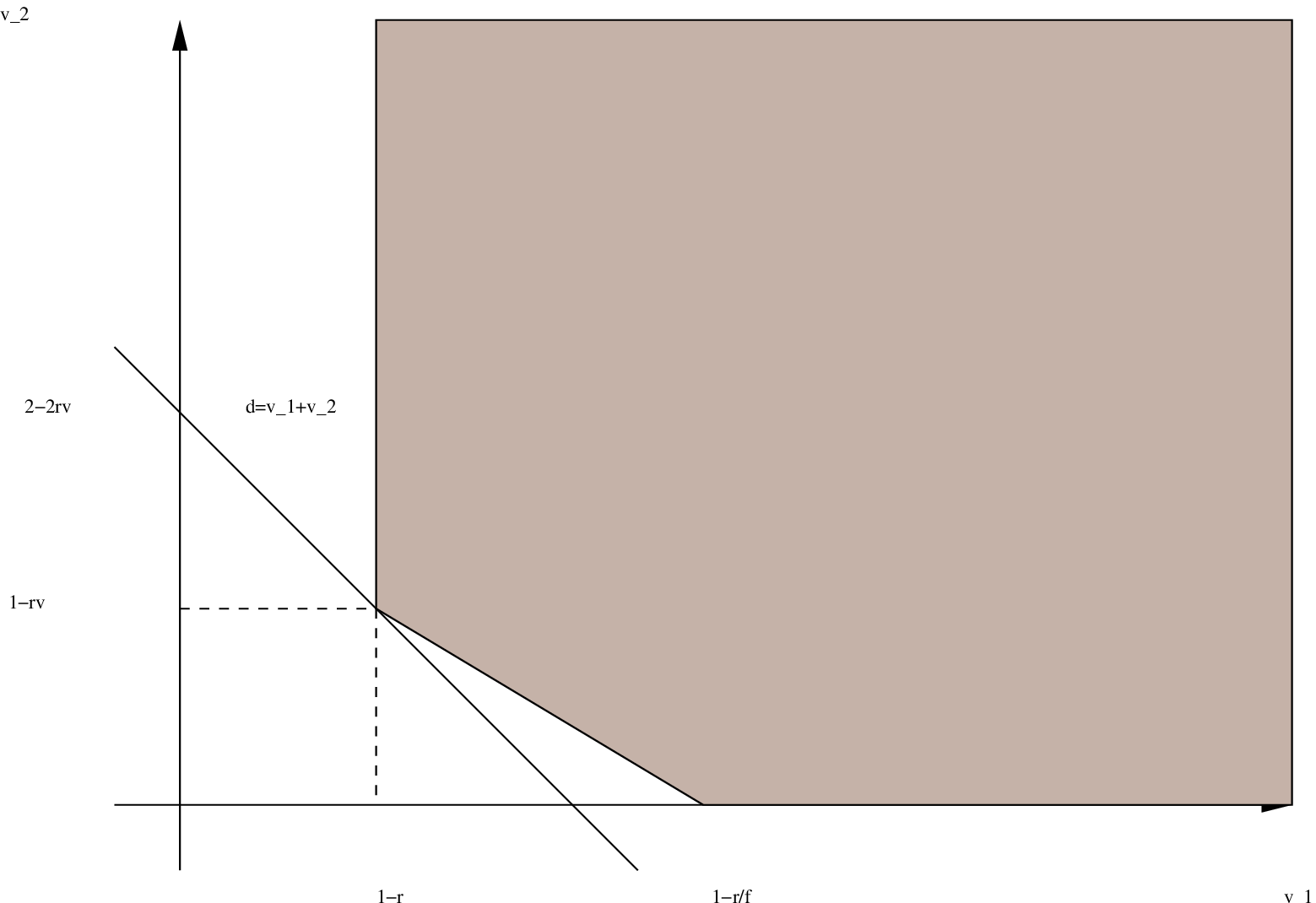}{Outage Region for the DDF protocol with a single
relay ($f\leq0.5$).}{3} {\footnotesize
 \psfrag{Outage}[]{$O^+$}
 \psfrag{v_1}[]{$v_1$}
 \psfrag{1-r/f}[]{$\frac{1-r}{f}$}
 \psfrag{1-r}[]{$1-r$}
 \psfrag{1-rv}[r]{$1-r$} \psfrag{d=v_1+v_2}[l]{$d=v_1+v_2$}
 \psfrag{2-2rv}[r]{$2(1-r)$} \psfrag{v_2}[r]{$v_2$}}

\putFrag{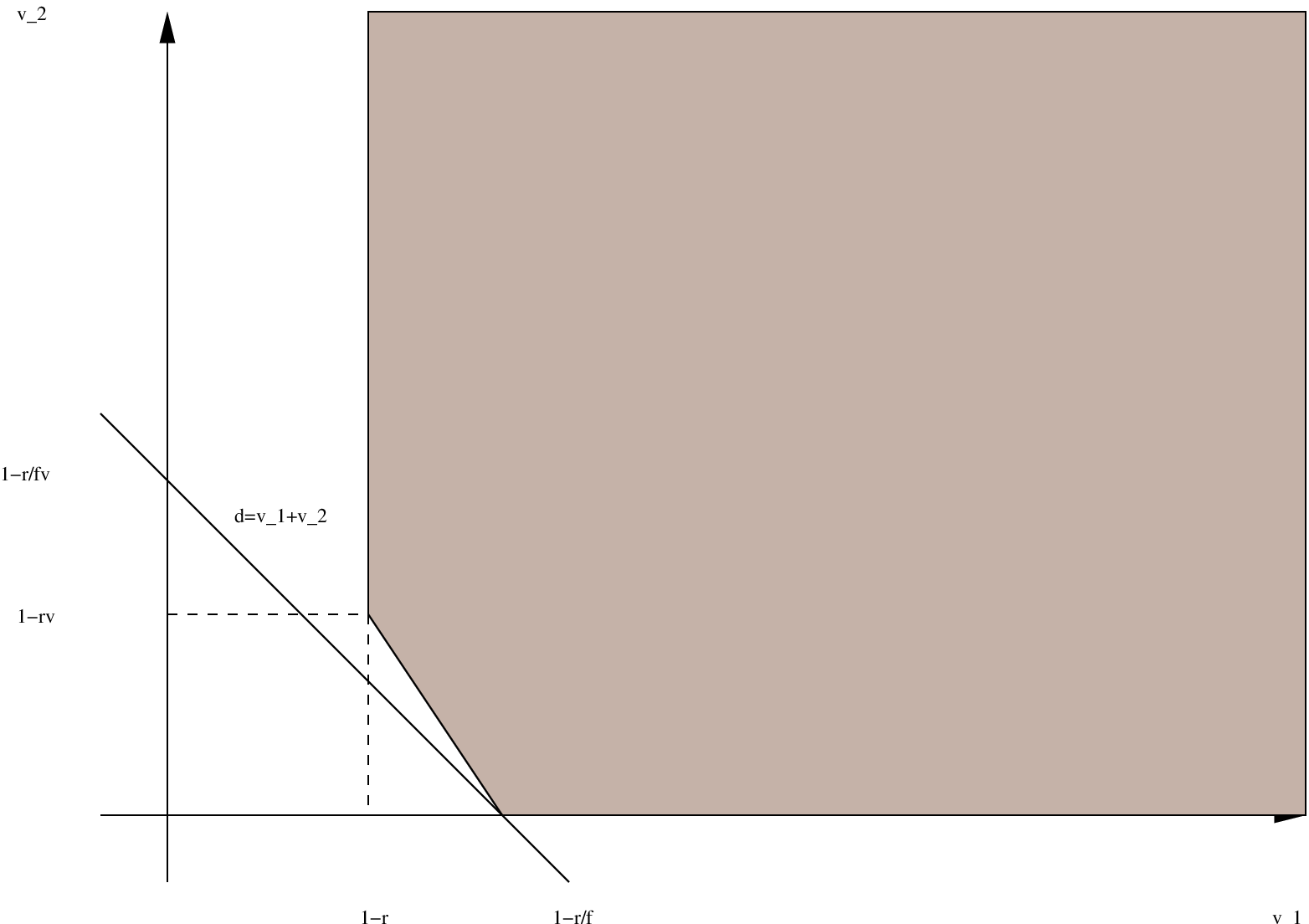}{Outage Region for the DDF protocol with a single
relay ($f>0.5$).}{3} {\footnotesize \psfrag{Outage}[]{$O^+$}
\psfrag{v_1}[]{$v_1$} \psfrag{1-r/f}[]{$\frac{1-r}{f}$}
\psfrag{1-r}[]{$1-r$} \psfrag{1-rv}[r]{$1-r$}
\psfrag{1-r/fv}[r]{$\frac{1-r}{f}$}
\psfrag{d=v_1+v_2}[l]{$d=v_1+v_2$}
\psfrag{1-r/1-f}[r]{$\frac{1-r}{1-f}$} \psfrag{v_2}[r]{$v_2$}}

\end{document}